\documentclass[twocolumn,tighten]{aastex701}

\begin{document}

\title{A Redshift-based Red Selection of Dusty Star-forming Galaxies}

\correspondingauthor{Amy J. Barger}
\email{barger@astro.wisc.edu}

\author[0000-0002-3306-1606]{A.~J.~Barger}
\affiliation{Department of Astronomy, University of Wisconsin-Madison,
475 N. Charter Street, Madison, WI 53706, USA}
\affiliation{Department of Physics and Astronomy, University of Hawaii,
2505 Correa Road, Honolulu, HI 96822, USA}
\affiliation{Institute for Astronomy, University of Hawaii, 2680 Woodlawn Drive,
Honolulu, HI 96822, USA}
\email{barger@astro.wisc.edu}

\author[0000-0002-6319-1575]{L.~L.~Cowie}
\affiliation{Institute for Astronomy, University of Hawaii,
2680 Woodlawn Drive, Honolulu, HI 96822, USA}
\email{cowie@ifa.hawaii.edu}

\author[0000-0003-4248-6128]{S.~J.~McKay}
\affiliation{Department of Physics, University of Wisconsin--Madison, 1150 University Avenue,
Madison, WI 53706, USA}
\email{sjmckay3@wisc.edu}

\author[0000-0002-8686-8737]{F.~E.~Bauer}
\affiliation{Instituto de Alta Investigaci{\'o}n, Universidad de Tarapac{\'a}, Casilla 7D, Arica, 1010000, Chile
}
\email{fbauer@astro.psu.edu}

\newcommand{\alma}{870\,$\mu$m}
\newcommand{\md}{\,M$_\odot$}
\newcommand{\sfr}{M$_\odot$\,yr$^{-1}$}
\newcommand{\fluxunits}{\,erg\,cm$^{-2}$\,s$^{-1}$}
\newcommand{\powerunits}{\,erg\,s$^{-1}$}
\newcommand{\esHz}{\,erg\,s$^{-1}$\,Hz$^{-1}$}

\begin{abstract}
We use JWST observations (1.5\,$\mu$m to 4.44\,$\mu$m),
together with complete ALMA observations (870\,$\mu$m and/or 1.2\,mm),
of the massive lensing cluster field A2744  to show that galaxies between $z=1.5$ and
$z=5.5$ with rest-frame red colors 
$f_J/f_V  > 3$ correspond to dusty star-forming galaxies
(DSFGs), little red dots (LRDs), and quiescent galaxies. The color selection picks out
34 of the 41 $>4.5\sigma$ ALMA sources in the field (83\%).
We find that the luminous 
[rest-frame $L_\nu(1.22\,\mu$m)$>3.6\times10^{29}$\,erg\,s$^{-1}$\,Hz$^{-1}$] red
sources are generally extended, while the less luminous red sources are almost all compact 
and correspond to the LRD population. We also find that
the great majority of the luminous, extended red sources are DSFGs based on the ALMA data, 
with a small admixture of quiescent galaxies at $z \lesssim 3$--4
that we identify based on their location in the rest-frame $U-V$ versus $V-J$ diagram. 
We do not detect any  LRDs or quiescent galaxies at the $>3\sigma$ level
in the ALMA images. Roughly 10\% of the DSFGs have high rest-frame X-ray luminosities
(here, $L(8$--28\,keV) $>2.5\times10^{43}$\powerunits)
and must be AGN dominated.
The DSFGs and quiescent galaxies
nearly all have $M_\ast>10^{10}$\md.
These massive galaxies become rare at $z>5$, paralleling 
the fall off in the number of detected DSFGs.
\end{abstract}

\section{Introduction}
\label{sec:intro}

Dusty star-forming galaxies (DSFGs), where the bulk of the light emerges
in the far-infrared (FIR), are a key population, providing much of 
the star formation over at least the redshift interval $z=2$--5 (e.g.,
\citealt{barger12,zavala21,sun25}). 
With their high star formation rates (SFRs), they may grow to become the bulk of 
the most massive galaxies at these redshifts (e.g., \citealt{barrufet25a,barrufet25b}).
Observationally, however, developing large samples of DSFGs with the accurate (arcsecond)
positions needed for spectroscopic follow-up and morphological classification has not been easy. 
Surveys with single-dish submillimeter/millimeter 
telescopes cover large areas but have poor spatial resolution and are
limited by confusion, while interferometric surveys with submillimeter/millimeter arrays
have the needed resolution but suffer from small-area coverage.
 
In order to avoid these problems, we need alternative ways of finding DSFGs.
Over the last decade, there has been considerable interest in using an
observed-frame red color selection to find large, highly complete 
samples of DSFGs with accurate positions 
(e.g., \citealt{wang12,wang19,barger23,barrufet23,barrufet25b,mckay24,mckay25}).
These methods are highly effective in finding DSFGs without
recourse to submillimeter/millimeter observations.
However, for fields with highly complete redshift information, we can improve on this approach
by using a rest-frame red color selection. This allows us to separate the different red 
populations, to map and compare them as a function of redshift, and to determine galaxy properties, 
such as NIR luminosity/stellar mass.

In this paper, we use the massive lensing cluster field A2744---with its vast redshift information
from JWST (e.g., \citealt{weaver24,price25,naidu25}), along with its wide-field ALMA data 
(\citealt{fujimoto25} and the present work)---to develop a red selection using 
rest-frame colors. By comparing with 
the ALMA data, we show that a rest-frame red color ($f_J/f_V>3$), together with a high
near-infrared (NIR) luminosity criterion ($L_\nu(1.22\,\mu$m)$>3.6\times10^{29}$\,erg\,s$^{-1}$\,Hz$^{-1}$),
is very effective in finding DSFGs. 

Our rest-frame red selection ($f_J/f_V>3$) also 
captures other interesting high-redshift galaxies, such as the ``little red dots"
(LRDs), which are currently suspected
to be high-redshift active galactic nuclei (AGNs)
\citep[e.g.,][]{matthee24,labbe25,hainline25,kocevski25,taylor25}.
These objects, which become more abundant at higher redshifts than
DSFGs, are readily separated from the DSFGs by their compactness and their comparatively low 
NIR to mid-infrared (MIR) luminosities/stellar masses.

At redshifts at or below $z\sim3$--4, our rest-frame red selection begins to include  
small numbers of quiescent galaxies, whose red colors are produced by
the sharp drop in flux at 4000~\AA\ (the Balmer break) due to their older stellar populations.
We can separate these galaxies from the DSFGs using their color-color properties
(e.g., \citealt{williams09,valentino23}). 

Finally, A2744 has deep Chandra
X-ray data \citep{bogdan24}, which we can use to identify luminous X-ray sources
that are dominated by AGNs.

The paper is structured as follows. In Section~2, we describe the JWST, ALMA, X-ray,
and redshift data that we use in our analysis. We also determine the demagnified rest-frame
fluxes and luminosities and give our adopted conversion to stellar mass.
In Section~3, we describe our red selection and the separation of the different populations
that comprise it. 
In Section~4, we present our analysis of the redshift evolution of the different red populations. 
Finally, in Section~5, we summarize our results.

We assume a standard cosmology of $H_0=70.5$~km~s$^{-1}$~Mpc$^{-1}$,
$\Omega_{\rm M}=0.27$, and $\Omega_\Lambda=0.73$ \citep{larson11}. In calculating
masses and SFRs, we assume a \cite{kroupa01} initial mass function.

\section{Data}
\label{sec:data}

\subsection {JWST}
\label{subsec:JWST}
The A2744 field has JWST observations from several surveys, including the GLASS-A2744 \citep{treu22} 
and UNCOVER \citep{bezanson24} programs.  Here we use the DR2 photometric catalog
from \citet{weaver24}, which contains additional data products provided 
by the UNCOVER team, including lensing magnifications, photometric and 
spectroscopic redshifts, and rest-frame HST and JWST fluxes in many bands. 
The provided lensing magnifications were derived from the lensing model of \citet{furtak23b}, 
which is highly constrained using HST and JWST multiply-imaged sources. 

We restricted to sources that lie in the ALMA covered area (see
Section~\ref{submm}) with signal-to-noise (S/N) above 20 in the
F444W band. This corresponds roughly to a lower limit
of 0.01\,$\mu$Jy for the F444W fluxes, and the median value is
$0.107$\,$\mu$Jy. The $1\sigma$ errors are as low as $0.005\,\mu$Jy.
We visually inspected each of the sources, and we removed any
blends and bad sources. This left us with a final sample of 5819 sources, which
we refer to as {\em our JWST sample}.

We measured the source sizes using 2D Gaussian fits to the F444W images, 
defining the FWHM as the geometric mean of the FWHMs in the two axes.
We consider sources with mean FWHM (hereafter, FWHM)
$<0\farcs172$ to be compact, where we chose the value to
match \cite{greene24}, who defined sources to be compact
if the ratio of the fluxes in $4\farcs0$ to $2\farcs0$ diameter
apertures is $<1.7$. This criterion should
select all sources that are small compared to the F444W 
point spread function (PSF) FWHM of $0\farcs148$.

\begin{figure}
\includegraphics[width=8.3cm,angle=0]{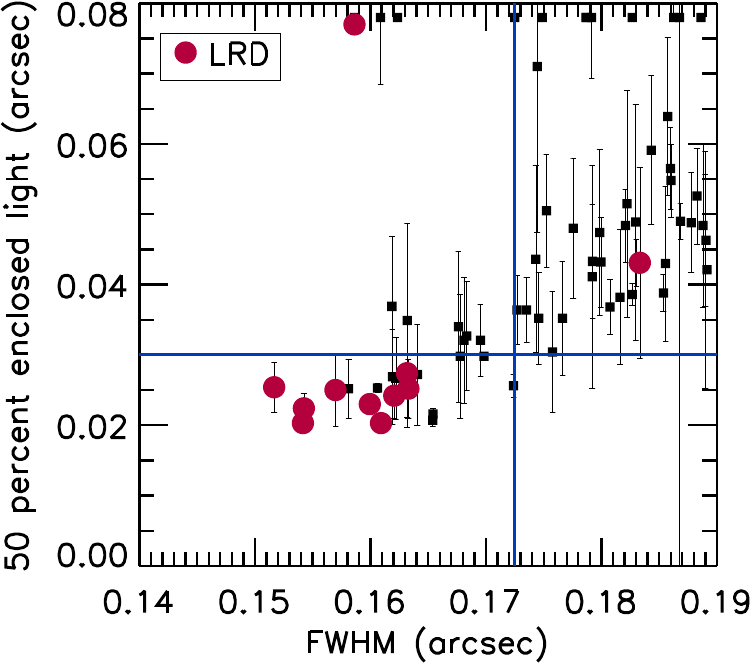}
\caption{
Comparison of the FWHM from the Gaussian fits with the
50\% enclosed light radius from the pysersic procedure of
\citet{pasha23}. The latter has the PSF deconvolved. The vertical
blue line shows our compact galaxy selection using FWHM, and the 
horizontal blue line a similar selection using the pysersic results. 
We show 68\% confidence ranges
for the pysersic results and place points that would lie
off the top of the figure at nominal values of $0\farcs077$.
We have not corrected either of these measures for lensing magnification.
Published LRDs are shown with red circles. 
\label{fwhm_comp}
}
\end{figure}

The Gaussian fit gives a robust measure of the sizes of the compact
sources we are interested in here, but we also fitted the sources
using the pysersic procedure of \citet{pasha23}. This allows us to determine the
50\% enclosed light radius with the PSF deconvolved.
We compare the two measures in Figure~\ref{fwhm_comp}, where
we also mark the published LRDs with red dots. In general,
the two measures agree well in the selection of compact sources
(blue lines show our compact galaxy selection for the two measures),
which is what we are interested in.
(Note that \citealt{miller26} present a detailed analysis of the size-mass relation from 
$0.5<z<8$ and $0.5<z<3$, for, respectively, star-forming and quiescent 
galaxies in A2744 using a morphology catalog based on 20-band JWST 
data from \citealt{zhang26}.)

However, in certain cases, the pysersic fits produce incorrect large values of the 
radii for blended or faint sources, as can be seen from Figure~\ref{fwhm_comp}.
For this reason, we prefer to use the Gaussian fits for our analysis.

\begin{figure*}
\includegraphics[width=8.3cm,angle=0]{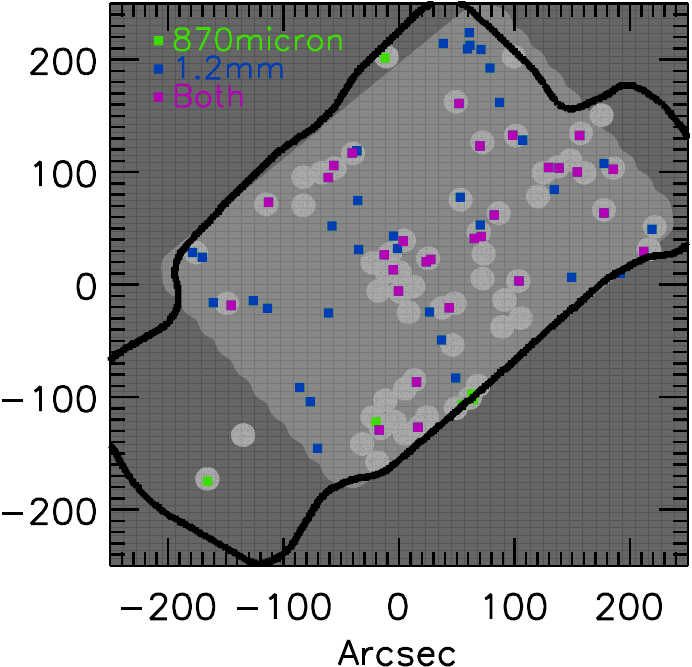}
\includegraphics[width=8.3cm,angle=0]{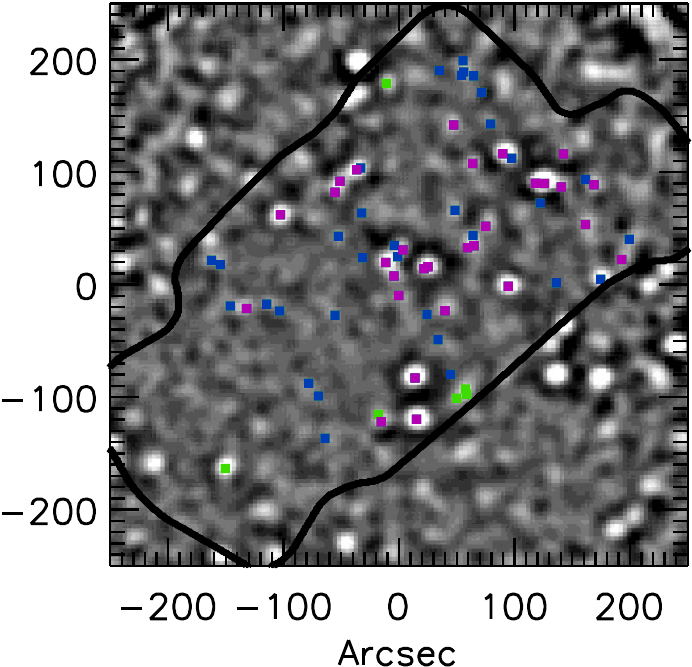}
\caption{ALMA observations in the JWST UNCOVER (black contour) area.
(Left) The shading shows areas covered by either DUALZ ALMA 1.2\,mm (light gray)
or this work's ALMA \alma\ (medium  gray). The pixel size
on the image is $0\farcs467$, and the total covered area is
29\,arcmin$^2$. Sources detected only at \alma\
are shown in green, only at 1.2\,mm in blue, and at both in purple.
(Right) The ALMA sources overlaid on the SCUBA-2 850\,$\mu$m image of the field with the
same color coding.
\label{schematic}
}
\end{figure*}

\subsection {Submillimeter/Millimeter}
\label{submm}
The SCUBA-2 data in A2744 were presented in \citet{cowie22} and discussed further in \citet{barger23} and \citet{mckay24}. 
The SCUBA-2 images are centered on the inner cluster region and have minimum 1$\sigma$ rms noise values at 
450\,$\mu$m and 850\,$\mu$m of 2.8\,mJy and 0.26\,mJy, respectively.

In addition to the previous ALMA coverage of the central A2744 cluster region from the 1.2\,mm ALMA Lensing Cluster Survey (ALCS; \citealt{fujimoto23a}) and the 1.1\,mm ALMA Frontier Fields Survey (AFFS; \citealt{gonzalez-lopez17, munoz-arancibia23}) surveys, \citet{fujimoto25} recently published the Deep UNCOVER-ALMA Legacy High-Z (DUALZ) survey data and catalog, which covered a 24\,arcmin$^2$ region corresponding to the UNCOVER 
NIRCam mosaic with 1.2\,mm ALMA imaging down to a minimum rms value of 32.7\,$\mu$Jy. 
We use the data products from that survey, since they include the previous ALMA data in the field.

Our own ALMA observations (Project 2023.1.00468.S) were carried out between April and August 2024
and consisted of 62 pointings toward $>4\sigma$ SCUBA-2 850\,$\mu$m sources in A2744.
The observations were performed in the C-3 configuration, giving angular resolutions 
of $\theta\sim0\farcs4$--$0\farcs5$ for a central frequency of 869\,$\mu$m.
We calibrated and processed our data using the standard ALMA data reduction pipeline with {\sc casa} version 6.5.4.9.
We generated cleaned data cubes for each pointing using {\tt robust = 0.5} and collapsed all spectral channels to produce 
continuum images.
We cleaned these images using the standard parameters, with a 2$\sigma$ clean threshold and automasking enabled. 
The final continuum images have typical (clean) beam sizes of  $0\farcs52 \times 0\farcs41$.
The typical $\sigma_{\rm rms}$ per pointing is 90~$\mu$Jy\,beam$^{-1}$.
To capture extended flux, we also produced $uv$-tapered versions of the continuum images using a 
1$\farcs$0 taper and the standard {\sc casa} reduction parameters.
The average $\sigma_{\rm rms}$ in the tapered images is $\sim$113~$\mu$Jy\,beam$^{-1}$.

We performed our source detection and flux measurement procedure on 
the cleaned images (untapered and tapered) using the {\tt pyBDSF} software
 \citep{mohanrafferty15}. {\tt pyBDSF} finds pixels above a set peak S/N threshold, 
$s_p$, and then identifies islands of emission above an ``island threshold,'' $s_i$.
It then fits Gaussians to each identified island and groups these into sources. In our case, since we expect only moderately extended sources, we set {\tt group\_by\_isl = True}, which groups all sources together within a single island.
Errors on the individual pixels and on the Gaussian-derived peak and total fluxes are computed by {\tt pyBDSF} using an rms map generated from the cleaned images.

We performed the source detection on the primary-beam-uncorrected images (since these have uniform noise characteristics) 
and corrected the fluxes and errors for primary beam attenuation afterwards.
We restricted our search to the area with primary beam response $>0.5$; i.e., the half-power beam width. 
For ALMA at \alma, this corresponds to a radius of 8$\farcs$75.

To determine our S/N threshold, we produced a histogram of all the pixels combined from all 80 ALMA pointings 
(uncorrected for the primary beam) and fit a Gaussian to the distribution.
For the untapered images, the best-fit Gaussian has $\sigma=87~\mu$Jy\,beam$^{-1}$.
At a threshold of $\sim4.5\sigma$, or 392 $\mu$Jy\,beam$^{-1}$, the expected number of false positives 
becomes $< 1$.  (For sources identified in other ways, we may use a lower threshold; in later sections,
we consider such sources measured at a $3\sigma$ level in either the \alma\ or DUALZ images.)
Thus, we set our direct detection S/N to $s_p=4.5$ in our {\tt pyBDSF} catalog construction procedure.
Although the noise characteristics are not identical across the different maps, they vary only slightly, and our catalog 
is not significantly affected by using a global peak threshold.
For the tapered images, the best-fit Gaussian has $\sigma=110~\mu$Jy\,beam$^{-1}$, and the number of false positives 
goes to $< 1$ at a threshold of $\approx 4.0\sigma$. Thus, we set our tapered S/N cutoff to $s_p=4.0$.

To further assess the purity of our resulting catalog, we also ran our source extraction on the inverse of each image
(i.e., the cleaned images multiplied by -1.0).
For $s_p=4.5$, we find just two detections on the negative untapered images.
This is slightly higher than the estimate based on Gaussian statistics (0.3 for $>4.5\sigma$,
given the roughly 78,000 independent beams in the dataset); however, it shows that our purity at this 
threshold is very high ($p \equiv 1 - N_{\rm neg}/N_{\rm pos} = 0.97$).

Our final ALMA \alma\ catalog consists of 38 $>4.5\sigma$ sources.
Some appear partially resolved in the untapered images at our resolution of $\sim 0\farcs5$. 
We therefore measured the fluxes for these sources using {\tt pyBDSF}'s 2D Gaussian fits to each source
in both the tapered and untapered images.
By comparing these various flux measurements to the SCUBA-2 850~$\mu$m flux 
at each source position, we determined that the total flux in 
the tapered images does the best job of recovering the source flux (i.e., minimizing losses due to extended emission).
Thus, we take the total fluxes in the tapered images as the \alma\ fluxes, 
except for the sources not detected at $>4.0\sigma$ in the tapered images, for which we instead use the total 
fluxes measured in the untapered images.

In the ALMA covered area of the JWST image,
68 sources are detected at $>4.5\sigma$
in at least one of the ALMA \alma\ and 1.2\,mm catalogs.
In Figure~\ref{schematic}, we show these sources (squares) marked on the
ALMA covered area (the contiguous DUALZ observed
area is shown in light gray, and this work's \alma\ observed areas in medium gray) (left panel)
and on the SCUBA-2 image (right panel).

In Table~\ref{almatable}, we give the ALMA total fluxes and uncertainties at both \alma\ and 
1.2\,mm for the 68 sources, uncorrected for magnification.
We also provide the magnifications from \citet{weaver24}. The
median magnification for the 68 sources is 1.5, and the maximum magnification is 4.6.
Finally, we give the JWST
F444W fluxes from \citet{weaver24}, but not the uncertainties, since these are so small.
Sources where there is
contamination or blending are marked with `?'. We measured source~3 directly from 
the F444W image, since there was not a measured value in \citet{weaver24}. 
We mark two sources with no F444W counterparts as `blank'. 
Both of these sources are at the very low S/N end of the 1.2\,mm data and may be spurious.

\subsection {X-ray}

\begin{figure}
\begin{center}
\vskip 0.4cm
\includegraphics[width=8cm,angle=0]{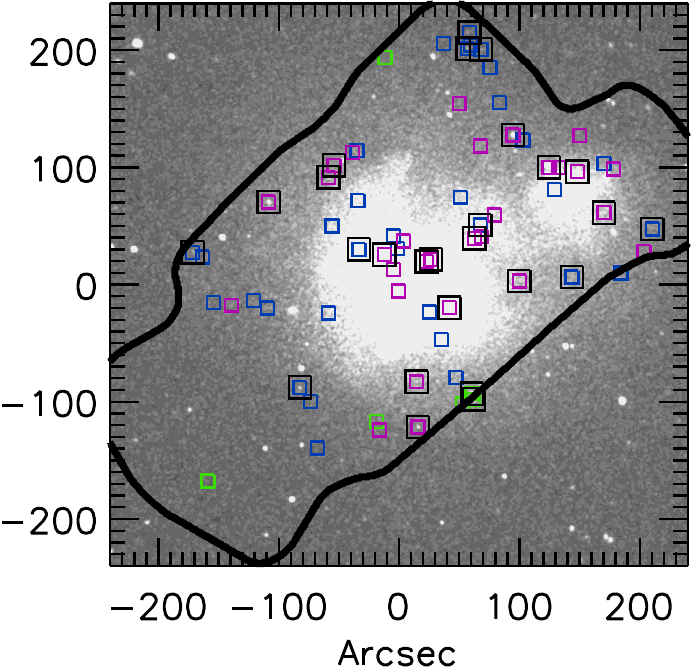}
\includegraphics[width=8cm,angle=0]{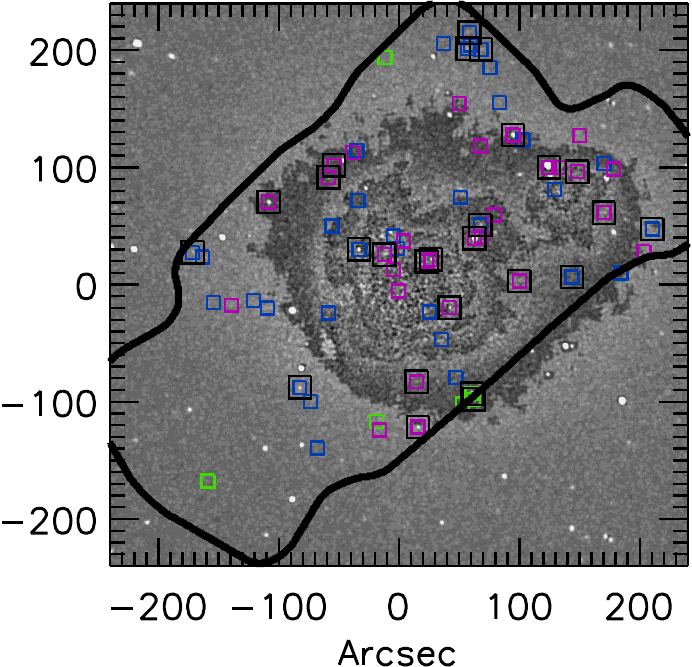}
\caption{The ALMA sources in the UNCOVER region 
(open squares using the color notation of Figure~\ref{schematic})
overlaid on the 0.5--7 keV Chandra image. In the top panel,
we show the X-ray image, which emphasizes the impact of the cluster 
foreground emission.  In the lower panel, we have subtracted a 
$7\farcs5$ median smoothed image to show the point sources in the
region of the cluster atmosphere. The black contour shows the JWST UNCOVER  area.
\label{xray_schematic}
}
\end{center}
\end{figure}

We took the  ACIS observations of A2744 from the Chandra X-ray Observatory archive. 
The total exposure time is 2.1\,Ms, 
but eliminating high-background periods reduces this to just over 2\,Ms.
We first reprocessed the individual observations using the CIAO tool CHANDRA REPRO. 
We then used the CIAO tool SRCFLUX to measure the X-ray fluxes at the
positions of the 68 sources in our JWST sample with $>4.5\sigma$ ALMA detections.
In Table~\ref{almatable}, we give the measured fluxes in the
soft (0.5--1.2\,keV), medium (1.2--2\,keV), and hard (2--7\,keV) bands for these sources.
Because of the increased cluster contamination
at the softer energies, we do not do any X-ray spectral distribution analyses.

We find that just over half (37) of the 68 sources are detected at the 90\% confidence level
in one or other of the bands. For comparison, in the deeper 7\,Ms CDF-S field,
\citet{barger19} found that two-thirds of the ALMA sample were
X-ray detected at $>3\sigma$.
As with previous studies in other fields (e.g., \citealt{ueda18,cowie18}), we find that
the fraction that are X-ray detected is higher 
when we restrict to the more luminous ($f_{870\,\mu{\rm m}} > 3$\,mJy) sources.

\begin{figure}
\begin{center}
\vskip 0.4cm
\includegraphics[width=8cm,angle=0]{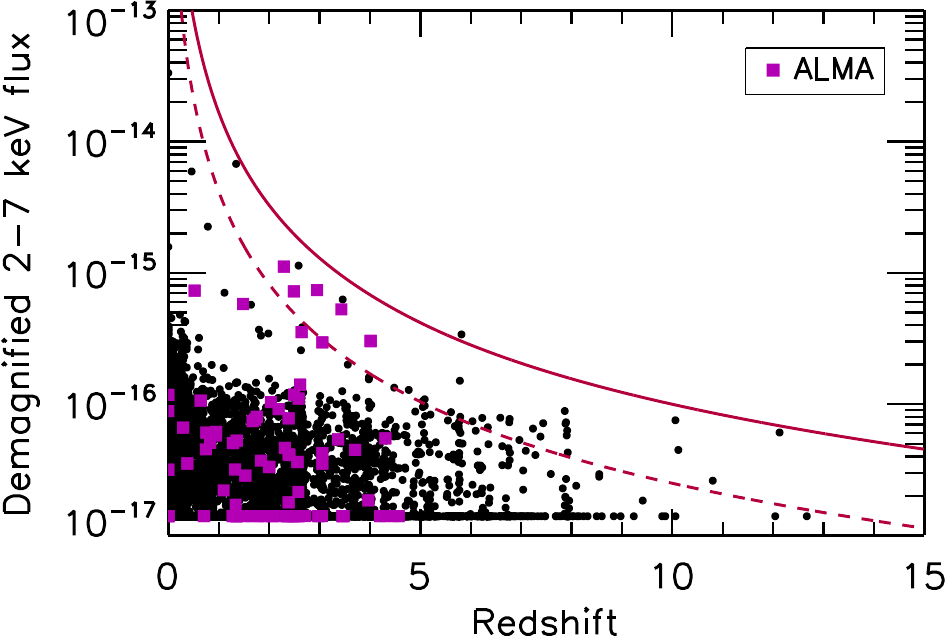}
\caption{Demagnified X-ray fluxes for our JWST sample vs.
redshift. The ALMA sources are marked in purple,
and faint X-ray sources are shown at $1.4 \times 10^{-17}$\fluxunits.
The red curves show a rest-frame 8--28\,keV luminosity of $10^{44}$\powerunits\
(solid) and  $2.5 \times 10^{43}$\powerunits\ (dashed).
\label{xray_sources}
}
\end{center}
\end{figure}

We computed the rest-frame luminosities at $z=3$ in order to match 
approximately the median redshift of our JWST sample.
Three of the 68 sources have demagnified rest-frame 8--28\,keV luminosities
(computed with Equation~3 of \citealt{cowie18}) $>2.5\times10^{43}$\powerunits,
which we will hereafter refer to as {\em bright X-ray sources}.
These sources must be dominated by AGN contributions rather than star formation.

We next formed exposure-weighted images using the CIAO4.17 tools. We
restricted to the hard (2--7\,keV) and broad (0.5--7\,keV) energy
bands, because of the high soft X-ray backgrounds owing to the cluster
atmosphere (see upper panel of Figure~\ref{xray_schematic}). 
Even with this restriction, it is necessary
to subtract off a smoothed image to see sources in the central
region of the image (bottom panel of Figure~\ref{xray_schematic}).
We then computed the observed frame 2--7\,keV fluxes at the positions of our JWST sample from
$1''$ diameter aperture counts with a 
background computed in a $3''$ to $5''$ annulus. 
The $1''$ diameter aperture corresponds to a 70\% to 80\%
encircled energy in this band over the JWST footprint.
We used a single conversion of $1.87\times10^{-19}$\fluxunits\ to convert from
counts\,s$^{-1}$. We show the demagnified 2--7\,keV fluxes versus redshift in Figure~\ref{xray_sources}.

As an aside, we note that, consistent with  
the results of \citet{ananna24}, none of the LRDs 
detected in the field (\citealt{greene24,labbe25,kocevski25})
are strongly detected in X-rays, with
all of the demagnified 2--7\,keV fluxes
$<10^{-16}$\fluxunits.

\subsection {Spectroscopic Redshifts}
\label{sec:redshifts}
We updated the \citet{weaver24} cataloged redshifts with recently measured JWST
spectroscopic redshifts \citep{greene24,price25,naidu25}.
The redshifts from these papers are 
generally consistent, with only two sources having significant
but minor discrepancies. In these cases, we adopted the redshifts of \citet{price25}. 
We took a further 8 spectroscopic redshifts
from the DAWN archive and 32 spectroscopic redshifts
from Keck/MOSFIRE data obtained as part of the present program. 
Only 3 of the DAWN redshifts are at $z<5.5$.

Our JWST sample contains 1250 sources with spectroscopic redshifts.
We adopted the \citet{weaver24} photometric redshifts for the remaining sources.
We note that while these are mostly consistent with the spectroscopic redshifts, for some sources,
the photometric redshifts substantially disagree with the spectroscopic redshifts.
For the 638 sources with spectroscopic redshifts between $z=1.5$ and $z=5.5$,
69 sources have a photometric redshift that differs by more than 0.2 from the 
spectroscopic redshift. This suggests that $\sim10$\% of the
photometric redshifts may be problematic.

\begin{figure*}
\centering
\includegraphics[width=0.495\linewidth]{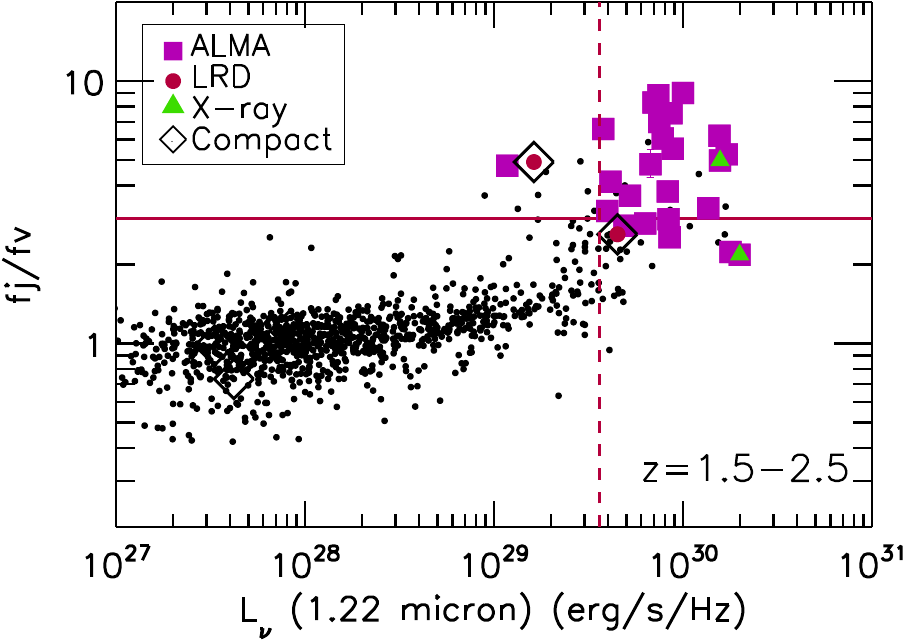}
\includegraphics[width=0.495\linewidth]{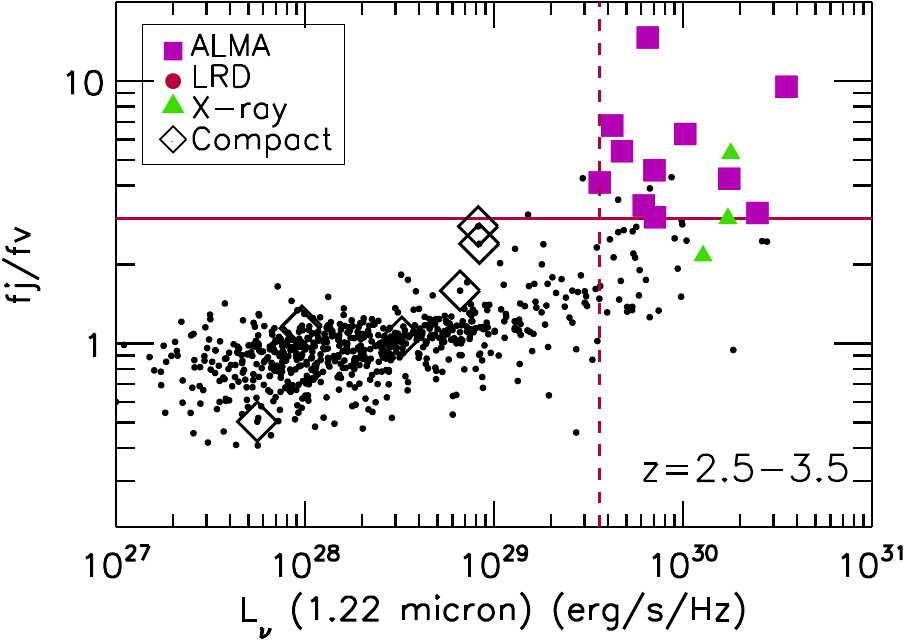}
\includegraphics[width=0.495\linewidth]{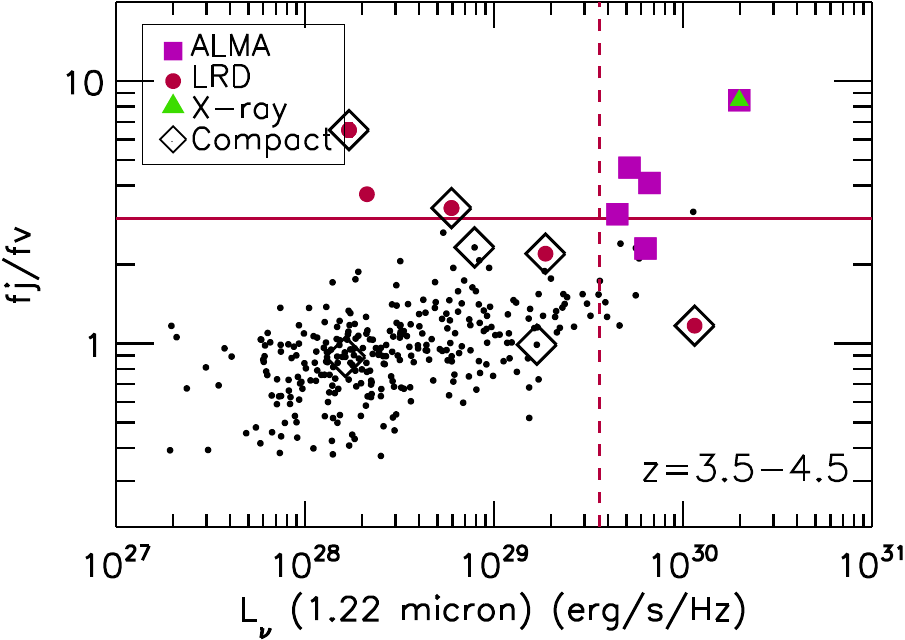}
\includegraphics[width=0.495\linewidth]{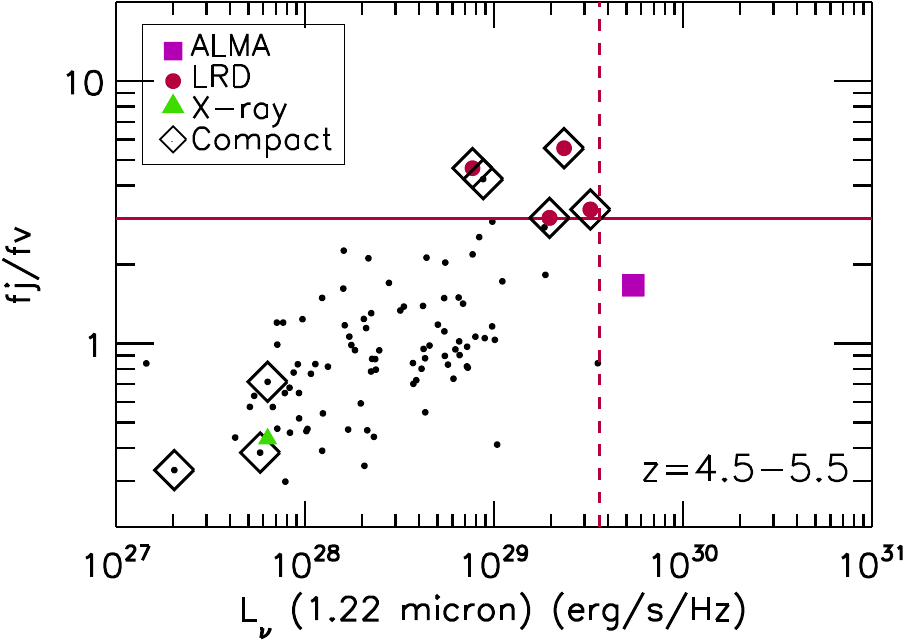}
\caption{Rest-frame color ($J=1.22~\mu$m and $V=0.55~\mu$m) vs.
demagnified rest-frame luminosity at 1.22\,$\mu$m
for our JWST sample with $L_\nu(1.22\,\mu{\rm m})>10^{27}$\esHz\ in four redshift intervals (black circles).
We overlay the sources with $>4.5\sigma$ ALMA detections (see Table~\ref{almatable}) 
with solid purple squares, 
LRDs (\citealt{greene24,labbe25,kocevski25}) with red circles, and
bright X-ray sources with green triangles.
We enclose compact sources with FWHM $<0\farcs172$ in open diamonds. 
The horizontal line denotes our red selection ($f_J/f_V$ $>3$), and the vertical line 
shows $L_\nu(1.22~\mu{\rm m})=3.6\times10^{29}$\esHz, which
corresponds to a stellar mass of $M_\star=10^{10}$\md.
The uncertainties are smaller than the symbol sizes.
\label{fig:colors}
}
\end{figure*}

\subsection{Rest-frame Colors and Luminosities}
\label{restframe}
We started from the rest-frame fluxes at $J$, $V$, and $U$ 
(1.22\,$\mu$m, 0.55\,$\mu$m, and 0.33\,$\mu$m)
given in \cite{weaver24}. The only exceptions are when
the new spectroscopic redshifts are substantially different than the
redshifts used by \cite{weaver24}; for these 150 sources, we
need to recompute the rest-frame fluxes. 
In order to minimize extrapolations in computing
the rest-frame flux at $J$, we restrict to sources with $z<5.5$.
We also restrict to sources with $z>1.5$, since the conversion
from 850\,$\mu$m flux to SFR only works well at these redshifts.
This leaves us with 2268 sources, of which 41 are in the $>4.5\sigma$ 
ALMA sample.

We then used the magnifications given in \cite{weaver24} to determine the demagnified 
rest-frame fluxes: $f_J$, $f_V$, and $f_U$. We did not recompute
the magnifications, since there is only a weak dependence 
on redshift. 

From $f_J$, we computed the demagnified
rest-frame luminosities at 1.22\,$\mu$m: $L_\nu$(1.22\,$\mu$m). This NIR luminosity
is a rough measure of the stellar mass, with 
%
%
\begin{equation}
M_\star=-0.410\times M_J + 0.85 \,,
\label{eq1}
\end{equation}
where $M_J$ is the absolute magnitude of $L_\nu$(1.22\,$\mu$m) (\citealt{nagaraj21}). 
Using this conversion, $L_\nu(1.22\,\mu{\rm m})=3.6\times10^{29}$\esHz\ corresponds to
a stellar mass of $M_\star=10^{10}$\md.

\cite{mckay25} measured masses for a number of their red selected sources in the
A2744 field using spectral energy distribution (SED) fitting with 
a \cite{calzetti00} extinction law. 
For their sources that are also included in our sample, we confirm that the masses 
are close, but the \cite{mckay25} values are better fit with the 0.85
in Equation~1 increased to 0.95. \cite{mckay25} noted the uncertainties
in the masses introduced by the choice of extinction law.
When appropriate, we use the mass conversion in Equation~\ref{eq1},
but given the uncertainty in the derived masses, we generally
prefer to use the directly measured luminosities.

\section{Red Galaxies}
\label{sec:redgal}
In Figure~\ref{fig:colors}, we plot $f_J/f_V$ versus $L_\nu(1.22\,\mu$m) for
our JWST sample in four redshift intervals: $z=1.5$--2.5, 2.5--3.5, 3.5--4.5, and 4.5--5.5,
with the $>4.5\sigma$ ALMA sources shown with purple squares.
As we discussed in Section~\ref{restframe}, there are 41 $>4.5\sigma$ ALMA sources 
in the redshift interval $z=1.5$--5.5 with well-defined counterparts in the JWST sample.
If we choose $f_J/f_V>3$ (horizontal line) as our red selection, then
34 (83\%) of these ALMA sources lie in this region.
Although a lower selection ratio would increase this completeness,
it would be at the expense of including more non-ALMA sources.
Thus, we define our red selection as $f_J/f_V>3$.
This selection also picks out most of the published LRDs (red circles),
but these tend to lie at lower $L_\nu(1.22\,\mu$m) than the ALMA sources (e.g., \citealt{ma25}).

If we choose $L_\nu$(1.22\,$\mu$m)~$>3.6\times 10^{29}$\esHz\ 
(vertical line) as our luminosity selection, then only one $>4.5\sigma$ ALMA source 
does not lie in this region (i.e., 1 out of 41).
This implies that 98\% of the ALMA sources correspond to high-mass
galaxies ($M_\ast>10^{10}$\md).

There are a total of 47 sources that have $L_\nu(1.22\,\mu$m)~$>3.6\times 10^{29}$\esHz\
and $f_J/f_V>3$, and hence lie in our 
selection quadrant. In addition to the 32 $>4.5\sigma$ ALMA sources that lie in the 
quadrant, there are a further 5 $>3\sigma$ ALMA sources there. 
We discuss these and the remaining 10 sources below.

\begin{figure}
\centering
\includegraphics[width=1.0\linewidth]{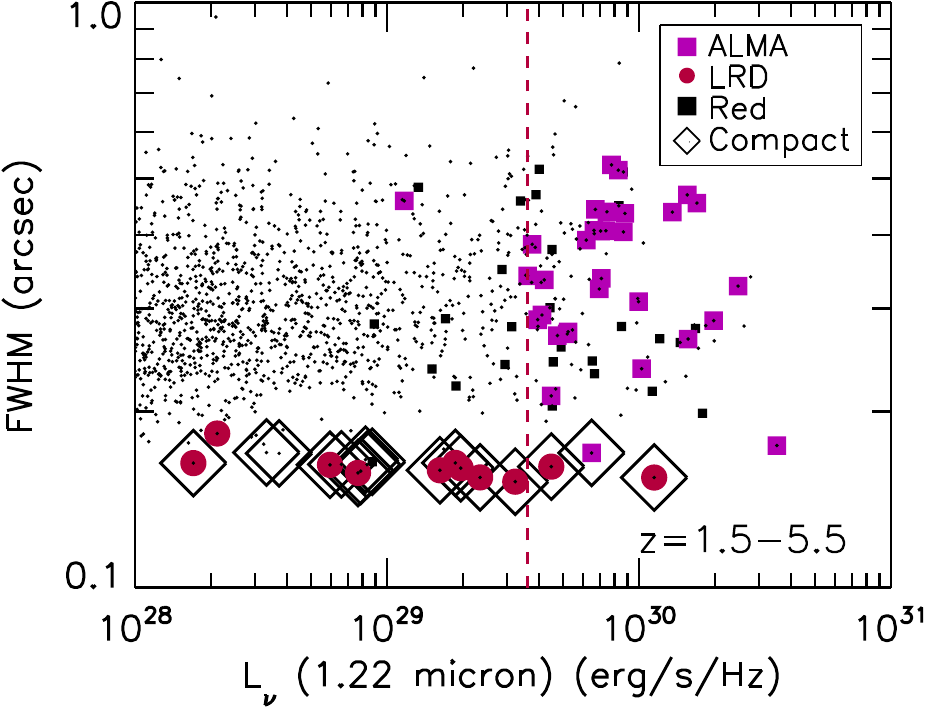}
\caption{FWHM vs. demagnified rest-frame luminosity at 1.22\,$\mu$m 
for our JWST sample with $L_\nu(1.22\,\mu{\rm m})>10^{28}$\esHz\ 
over the full redshift interval $z=1.5$--5.5 (small black dots).
We mark compact sources with FWHM $<0\farcs172$ with
large open diamonds.
We mark galaxies that satisfy our red selection ($f_J/f_V>3$) as follows:
sources with $>4.5\sigma$ ALMA detections (see Table~\ref{almatable}) 
with solid purple squares; LRDs (\citealt{greene24,labbe25,kocevski25}) 
with solid red circles; and other red  galaxies with solid black squares.
The vertical line shows 
$L_\nu(1.22~\mu{\rm m})=3.6\times10^{29}$\esHz, which
corresponds to a stellar mass of $M_\star=10^{10}$\md.
\label{ref_flj_wid}
}
\end{figure}

\begin{figure}
\begin{center}
\includegraphics[width=6.5cm,angle=0]{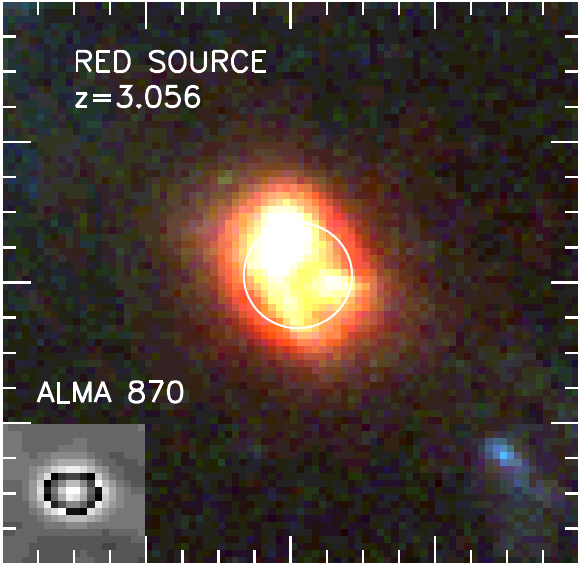}
\includegraphics[width=6.5cm,angle=0]{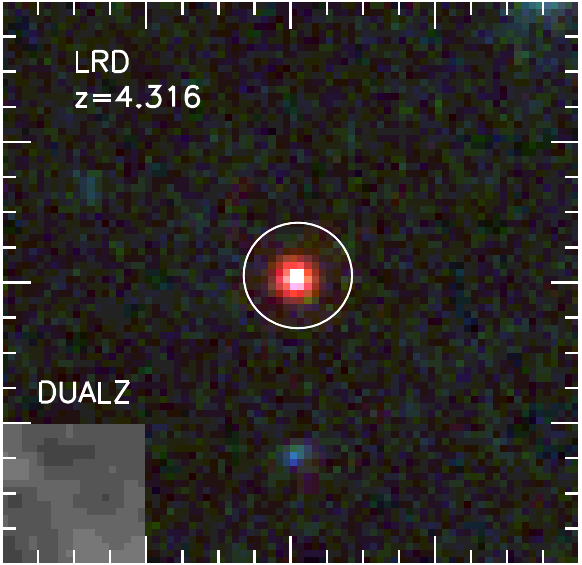}
\vskip -0.1cm
\caption{(Left) RGB image (F444W, F200W, and F150W)
of an example extended red source (white
circle of radius $0\farcs3$). The field size is $3\farcs2$.
The 870\,$\mu$m ALMA image of the same area is shown in the
inset at the lower left. The galaxy is strongly detected
with ALMA at both 870\,$\mu$m and 1.2\,mm. 
(Right) Same for an example compact red source, which
is an LRD. Here the inset is from the DUALZ image,
and there is no ALMA detection.
\label{samples}
}
\end{center}
\end{figure}

\subsection{Separating DSFGs from LRDs}
Our main goal here is to separate extended sources from compact ones.
A more detailed discussion of the sizes of DSFGs in the A2744 and GOODS-S fields
can be found in \cite{mckay25}.

In Figure~\ref{ref_flj_wid}, we plot FWHM versus $L_\nu(1.22\,\mu$m) over the full
redshift interval $z=1.5$--5.5. 
We do not demagnify the measured observed sizes, since these contain 
substantial contributions from the instrument PSF, and we are only
using these to distinguish between extended and compact sources.
From the figure, we see that there are only a very small number of compact sources;
there are 18 with FWHM $< 0\farcs172$. These include 10 of the
LRDs in the region, together with one ALMA source and one further red source.

Nearly all of the sources with $>4.5\sigma$ ALMA detections
(purple squares) are extended, with F444W widths
comparable to those of the general galaxy population.
In the redshift interval $z=3$--5, the mean size of the sources with
$>4.5\sigma$ ALMA detections
is $0\farcs27$, while for those without, it is
$0\farcs26$. At $z=2$--3, these numbers are,
respectively, $0\farcs34$ and $0\farcs31$. 

In Figure~\ref{samples}, we show examples of both an extended red source
and a compact red source.

Some of the red sources that are extended but
do not have $>4.5\sigma$ ALMA detections may be quiescent galaxies
(black squares). We discuss them in Section~\ref{sec:QG}.

\subsection{Separating DSFGs from Quiescent Galaxies}
\label{sec:QG}
The full-coverage ALMA data make it possible to examine
the color separation between DSFGs and quiescent galaxies. 

\subsubsection{Massive Populations}
\label{sec:massive}
\begin{figure}
\vskip 0.5cm
\centering
\includegraphics[width=0.9\linewidth]{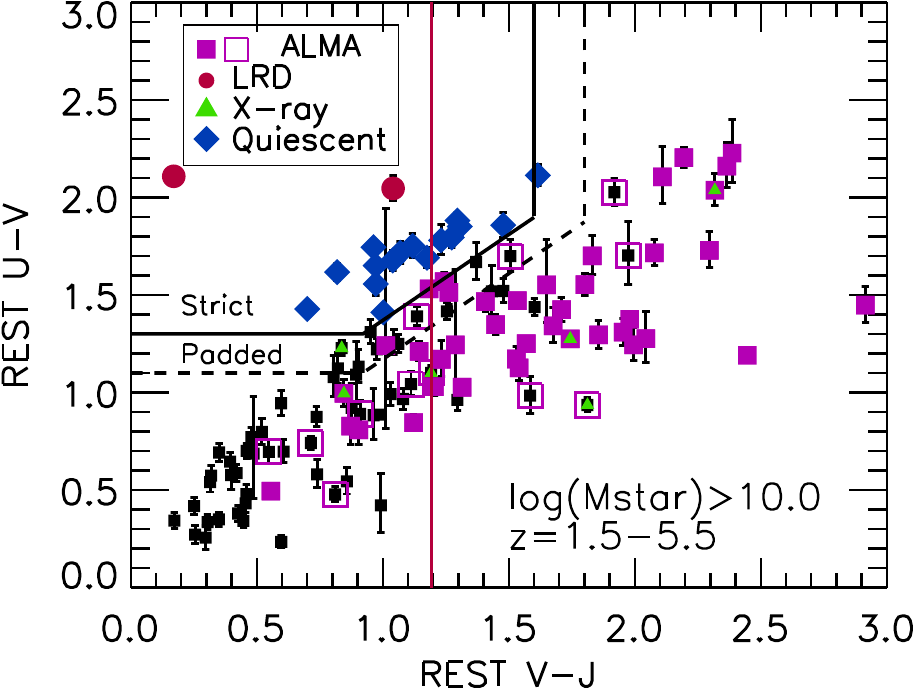}
\vskip 0.5cm
\includegraphics[width=0.9\linewidth]{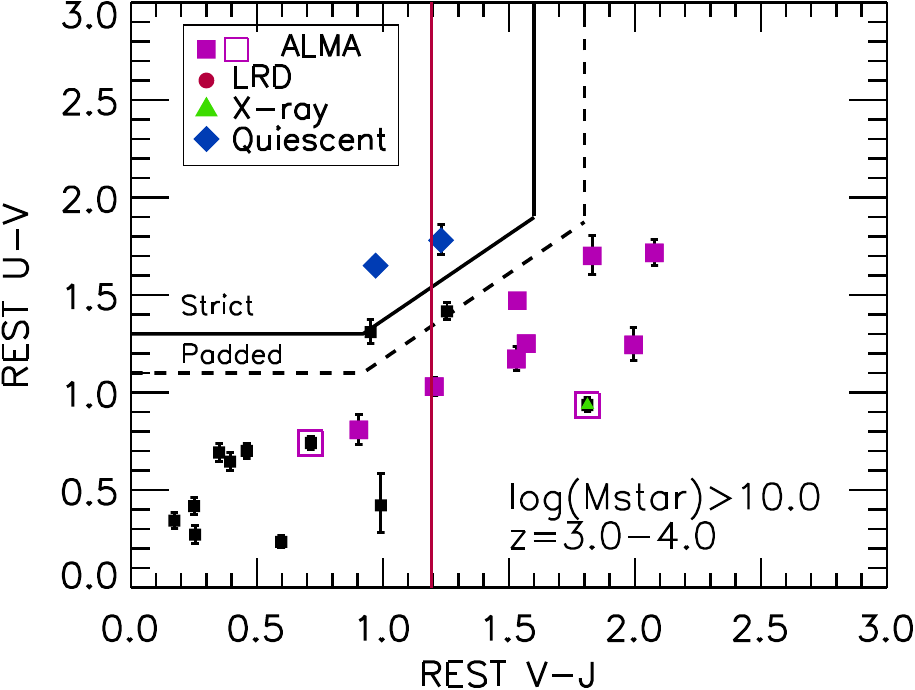}
\vskip 0.5cm
\includegraphics[width=0.9\linewidth]{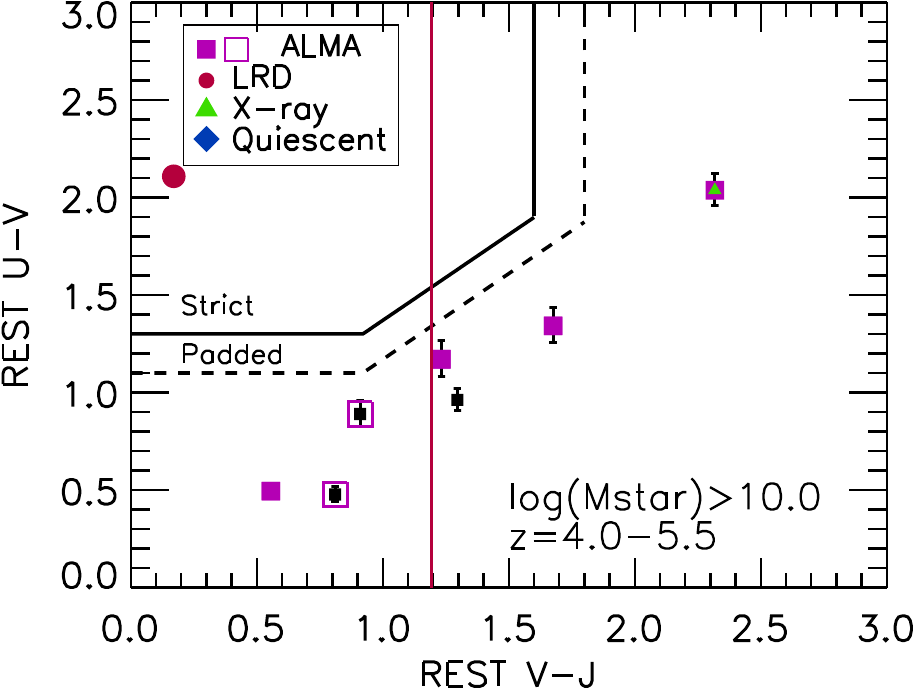}
\caption{Rest-frame $U-V$ vs. $V-J$ for our JWST sample 
with stellar masses $M_\ast>10^{10}$\md\ and that lie in
(top) the full redshift interval $z=1.5$--5.5, (middle) $z=3$--4 only, and (bottom) $z=4$--5.5 only.
We show the \cite{williams09} strict selection of quiescent galaxies with solid lines, 
and the \cite{valentino23} padded selection with dashed lines.
We show sources that satisfy the strict selection with blue diamonds,
except for luminous LRDs, which we  show as red circles.
We also show: ALMA detections---purple squares (solid for sources with $>4.5\sigma$ 
detections; open for $>3\sigma$);
bright X-ray sources---green triangles; remaining sample---black squares.
The vertical red line marks our red color selection 
(see Section~\ref{sec:redgal}), with red objects lying to the right.
\label{fig:color_select}
}
\end{figure}

In Figure~\ref{fig:color_select}, we plot rest-frame $U-V$ versus 
$V-J$ for our JWST sample with $M_\ast >10^{10}$\md\ over the
full redshift interval $z=1.5$--5.5, and for two higher redshift intervals
separately ($z=3$--4 and $z=4$--5.5).
We show the strict color selection
of \cite{williams09} with the solid lines, and the padded color selection
of \cite{valentino23} with the dashed lines. We show the sources lying in the
strict region as blue diamonds, except for the two luminous LRDs, which we 
show as red circles.

\begin{figure}
\centering
\includegraphics[width=1.05\linewidth]{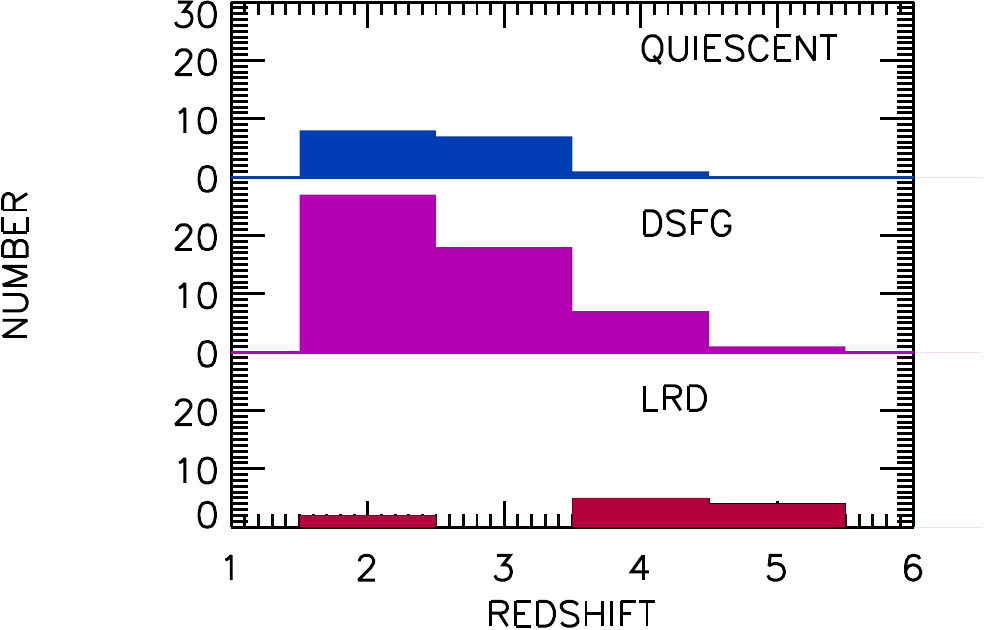}
\caption{Relative evolution of the numbers of candidate quiescent 
galaxies (blue), DSFGs (purple), and LRDs (red)
in the redshift interval $z=1.5$--5.5 vs. redshift.
For the quiescent galaxies and DSFGs, we only show 
those with stellar masses $M_\ast>10^{10}$~M$_\odot$. 
The DSFGs show all $>3\sigma$
ALMA sources, and the quiescent galaxies correspond
to the strict selection.
\label{num_evol}
}
\end{figure}

Of the 123 sources at $z=1.5$--5.5 (top panel of Figure~\ref{fig:color_select}), 
and excluding the two luminous LRDs,
there are 16 sources in the strict region. None of these 16 sources has
 a $>3\sigma$ ALMA detection. All of these sources are covered by the DUALZ
 imaging, and their mean 1.2\,mm flux is $0.059\pm0.040$\,mJy.
The padded region shows a roughly equal mixture of $>3\sigma$ ALMA detections 
and potentially quiescent galaxies (5 and 7, respectively). 

For the sources at $z=4$--5.5 (bottom panel of Figure~\ref{fig:color_select}), 
we see that there are no quiescent galaxies at these redshifts.
Moreover, nearly all of the massive galaxies here (6 out of 7) are DSFGs.

Although small numbers of quiescent galaxies appear at $z=3$--4
(middle panel of Figure~\ref{fig:color_select}) (two in the strict 
region, and two more in the padded region),
together with a number of blue galaxies ($U-V<0.75$), DSFGs remain the
predominant population, with 10 sources in this redshift interval.

We show this in histogram form in Figure~\ref{num_evol}, where we see that 
compared to the LRDs, which mostly lie at the very high end of the redshift range,
the DSFGs begin to enter at $z<5$ and the quiescent galaxies at $z<4$. We will 
consider what this implies for the formation and evolution of the massive galaxies 
in the Discussion.

As an aside, we also investigated alternate color selections, such as the $ugi$  method
of \cite{antwidanso23} and the NUV$-r-J$ selection of \cite{ilbert13}.
The selections of quiescent galaxies from these methods are broadly
consistent with the $UVJ$ method, but the separations are less clean, so we 
adopt the $UVJ$ method. 
This has the additional advantage that it is the method
most often used in the literature.

In the following, we will consider galaxies lying in the strict region
(excluding the two luminous LRDs)
plus galaxies lying in the padded region without $>3\sigma$ ALMA detections as 
possible quiescent galaxies.

\subsubsection{Lower Mass Populations}
\label{sec:lowermass}
\begin{figure}
\includegraphics[width=8cm]{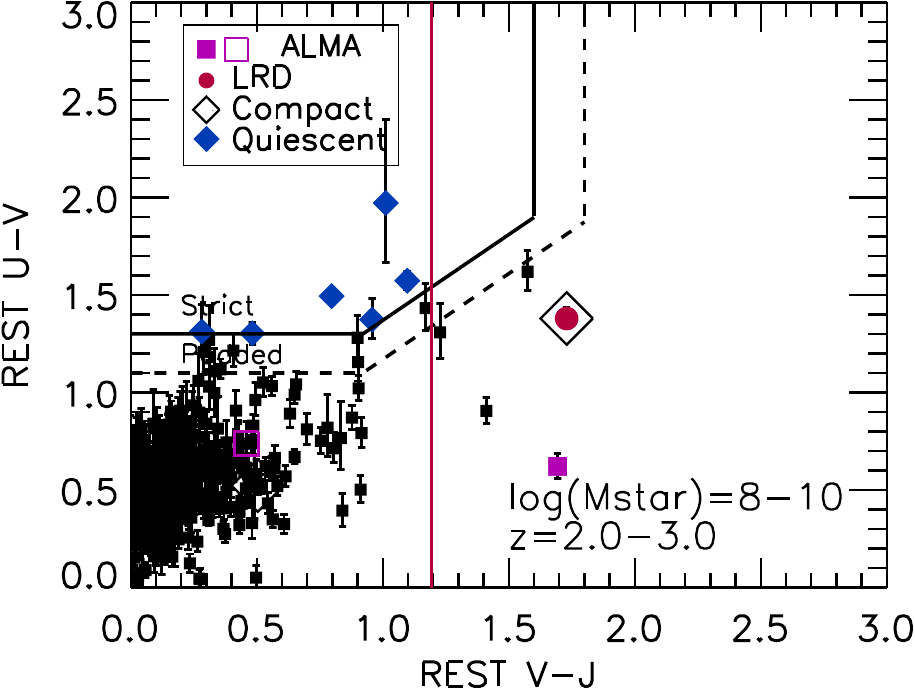}
\vskip 0.5cm
\includegraphics[width=8cm]{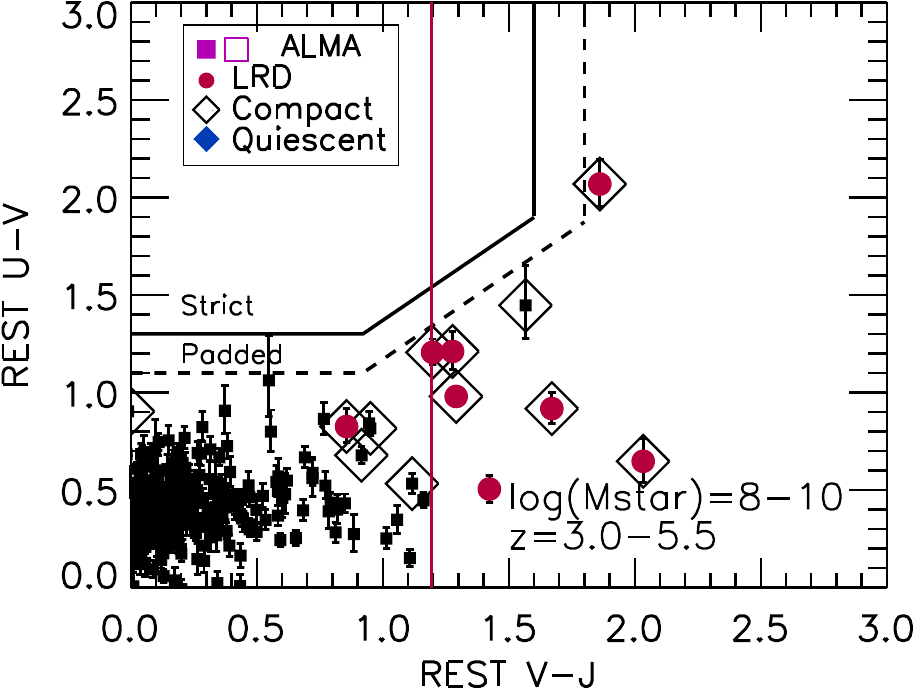}
\caption{Rest-frame $U-V$ vs. $V-J$ for our JWST sample
with stellar masses $M_\ast=10^{8}$--$10^{10}$\md\ and that lie in the
redshift intervals
(top) $z=2$--3 and (bottom) $z=3$--5.5.
We show the \cite{williams09} strict selection of quiescent galaxies with solid lines, 
and the \cite{valentino23} padded selection with dashed lines.
We show sources that satisfy the strict selection with blue diamonds.
The other symbols are ALMA detections---purple squares (solid for sources
with $>4.5\sigma$ detections; open for $>3\sigma$); 
LRDs---red circles; other compact sources with 
FWHM $<0\farcs172$---large open diamonds;
remaining sample---black squares.
The vertical red line marks our red color selection (see Section~\ref{sec:redgal}), 
with red objects lying to the right.
\label{fig:color_color_lowmass}
}
\end{figure}

In Figure~\ref{fig:color_color_lowmass}, we plot rest-frame $U-V$ versus 
$V-J$ for our JWST sample with $M_\ast=10^8$--$10^{10}$\md, separated 
into two redshift intervals: $z=2$--3 and $z=3$--5.5. 
At these lower stellar masses, 
more LRDs (red circles) and other compact sources (large open diamonds) begin to appear.
There are a total of 24 compact sources with FWHM $<0\farcs172$
in this mass range  between $z=2$ and $z=5.5$, 8 of which have been classified as LRDs.
We do not detect any of the 24 sources at the $>3\sigma$ level in the ALMA images.
Excluding the LRDs, there are 12 compact sources that are
blue with $V-J <0.75$.
Four compact sources in the $z=3$--5.5 range are redder 
and could be additional LRDs, including an extremely red source at
RA: $0^h$  $14^m$  $11.45^s$, Decl:  $-30^\circ$  $20'$  $1\farcs3$
with a spectroscopic redshift of $z=4.874$ from both \citet{price25} and \citet{naidu25}.
The prism spectrum of this source appears to show a weak underlying
broad line in H$\alpha$, which would be consistent with it being an LRD.
There are also no DSFGs or quiescent galaxies above $z=3$ at these masses.
However, there are six quiescent galaxies and one DSFG
in the low-mass interval at $z=2$--3.

\section{Discussion}
\label{discuss}
Studies like \cite{barger14} and \cite{cowie17} showed that there is a
close linear relation between the SFR and the 850\,$\mu$m flux
for bright ($>2$~mJy) DSFGs. 
\cite{mckay25} demonstrated that this relation 
extends to the faint ($<2$~mJy) DSFGs, as well,
by selecting a sample of red galaxies in the A2744 and CDF-S
fields using observed-frame colors and fluxes:
$f_{\rm F444W}/f_{\rm F150W} > 3.5$ and 
$f_{\rm F444W} > 1\,\mu$Jy.
By converting the integrated IR luminosities that they obtained from full galaxy SED fits with the
BAGPIPES code \citep{carnall18}, they computed magnification-corrected SFRs for their sample. 
(Note that BAGPIPES uses a fully Bayesian framework to constrain the posterior
likelihood distributions of the model parameters.)
They then performed a linear fit to the 234 sources in their sample, along with 1130
sources from five literature samples 
(the 99 sources in the main ALESS sample (\citealt{dacunha15}); the 707 sources in
the AS2UDS sample (\citealt{dudzeviciute20}); the 289
sources in the SCUBADive sample (\citealt{mckinney25});
and the 35 sources in the main ASPECS sample (\citealt{aravena20}),
weighted by the individual errors on the data points, to obtain
\begin{equation}
{\rm SFR} \, \,  [M_\odot \, {\rm yr}^{-1}] = (83\pm 11) \times f_{850~\mu{\rm m}} \,\, {\rm [mJy]}
\label{mckayeq}
\end{equation}
for the range 0.1--10 mJy (the error was derived from bootstrapping
the best-fit relation). It is important to stress that this relation is independent of redshift.

Although \cite{dudzeviciute20} claimed a stronger correlation of 
$f_{850~\mu{\rm m}}$ with cold dust mass using the AS2UDS sample alone, it is clear that
the linear $f_{850\,\mu{\rm m}}$--SFR relation of \cite{mckay25}
(see their Figure~11) provides a good fit across a wide range of submillimeter fluxes.
We note that previous determinations of this relation have been higher
(e.g., SFR=$134\times f_{850\,\mu{\rm m}}$ from \citealt{barger14}, when converted to a
\citealt{kroupa01} initial mass function, and SFR=$143\times f_{850\,\mu{\rm m}}$ from
\citealt{cowie17}), but this could reflect systematic differences in the way the SFRs are calculated
(i.e., direct IR luminosity conversions versus SFRs derived from SED codes).
The above differences are probably a good representation of the systematic uncertainties.

We find that
61 of the \cite{mckay25} sources also appear in our JWST sample in the redshift interval 
$z=1.5$--5.5. Thus, we adopted their SFRs for these 61 sources.
For the remaining sources in our JWST sample, we computed the SFRs using
Equation~\ref{mckayeq}, after correcting the fluxes for magnification.

To do this, we first  cross-correlated the ALMA data with the SCUBA-2 data
to estimate 850\,$\mu$m fluxes for the ALMA sources. 
For each of the sources from our ALMA program, we applied
a multiplicative correction of 1.13 to the total ALMA \alma\
flux. This correction factor is based on the mean ratio
of the SCUBA-2 850\,$\mu$m flux to the ALMA \alma\ flux for sources 
with ALMA \alma\ flux $>1.5$\,mJy. 
For each of the sources from the DUALZ program, we applied a multiplicative
correction of 2.55 to the total ALMA 1.2\,mm flux. This correction
factor is based on the mean ratio of the SCUBA-2 850\,$\mu$m flux to the 
ALMA 1.2\,mm flux for sources with ALMA 1.2\,mm flux $>0.5$\,mJy.
Next, we calculated the SFR from the $f_{850~\mu{\rm m}}$--SFR relation of
\citet{mckay25} (see Equation~\ref{mckayeq}).

Both our ALMA \alma\ fluxes and the DUALZ 1.2\,mm fluxes
give very similar demagnified SFR sensitivities, i.e., 
a $4\sigma$ of 26 \sfr\ and 30 \sfr, respectively, for a median
magnification of 1.62 and an rms of 0.11\,mJy (our ALMA \alma)
and 0.056\,mJy (wide DUALZ 1.2\,mm).
{\em In the following, we consider galaxies with SFRs $>30$ \sfr\
to be powerfully star-forming galaxies.} The remaining galaxies
could be quiescent, or they could have star formation below this level.

We find that in our sample, there are 59 powerfully star-forming galaxies 
in the redshift interval $z=1.5$--5.5. 
While 43 of these are also in \cite{mckay25},
the remaining 16 have single-flux SFRs and are roughly equally divided
between our ALMA data and the DUALZ data. Of the 59 galaxies, 31
have spectroscopic redshifts, and the rest have photometric redshifts.

\begin{figure}
\includegraphics[width=8cm,angle=0]{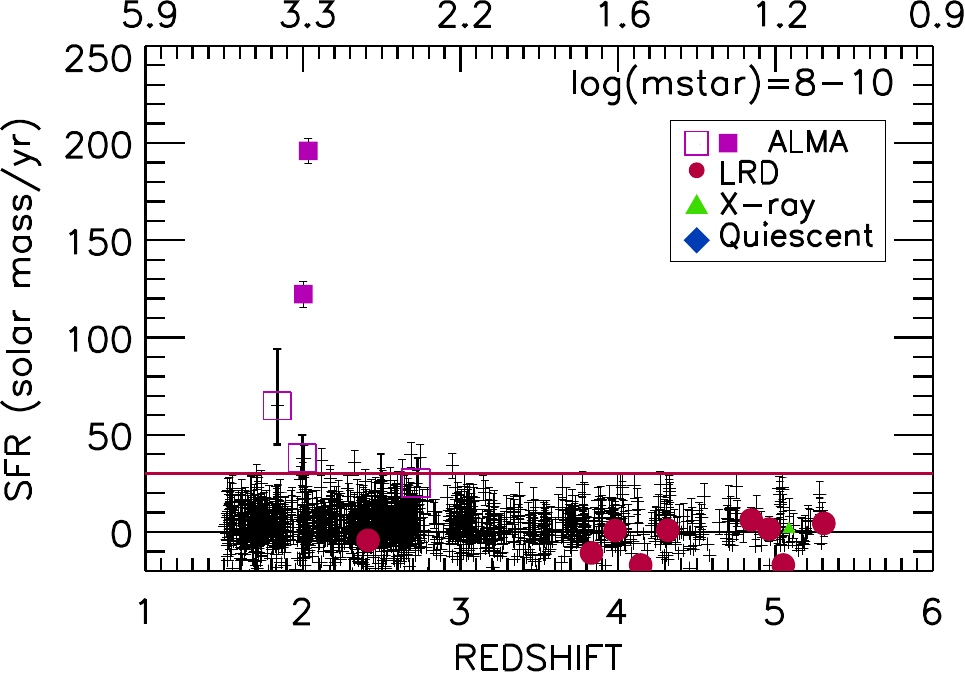}
\vskip 0.5cm
\includegraphics[width=8cm,angle=0]{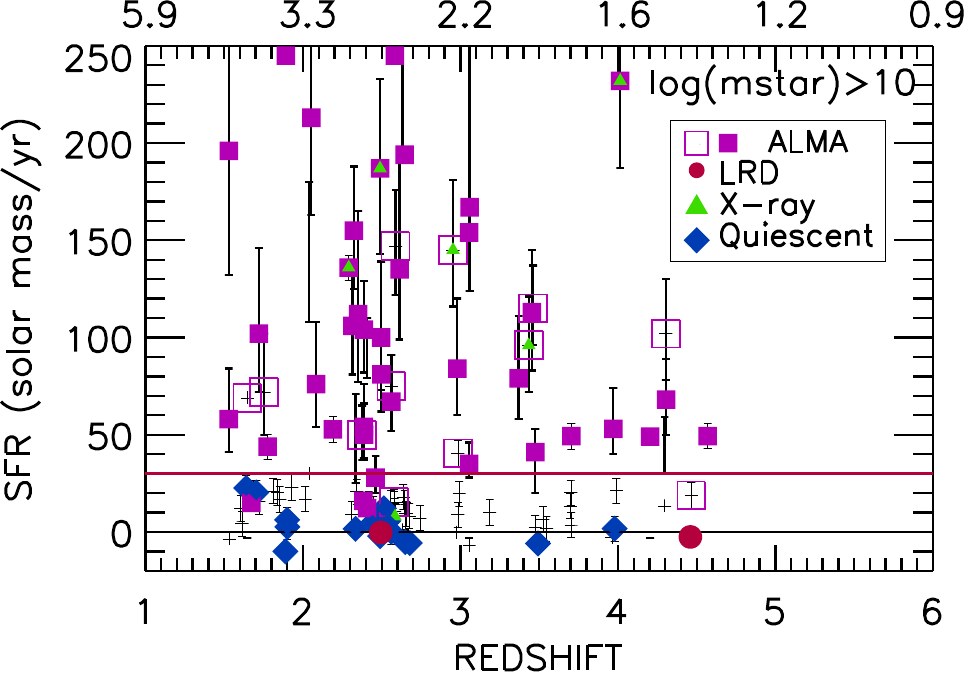}
\vskip 0.5cm
\includegraphics[width=8.2cm,angle=0]{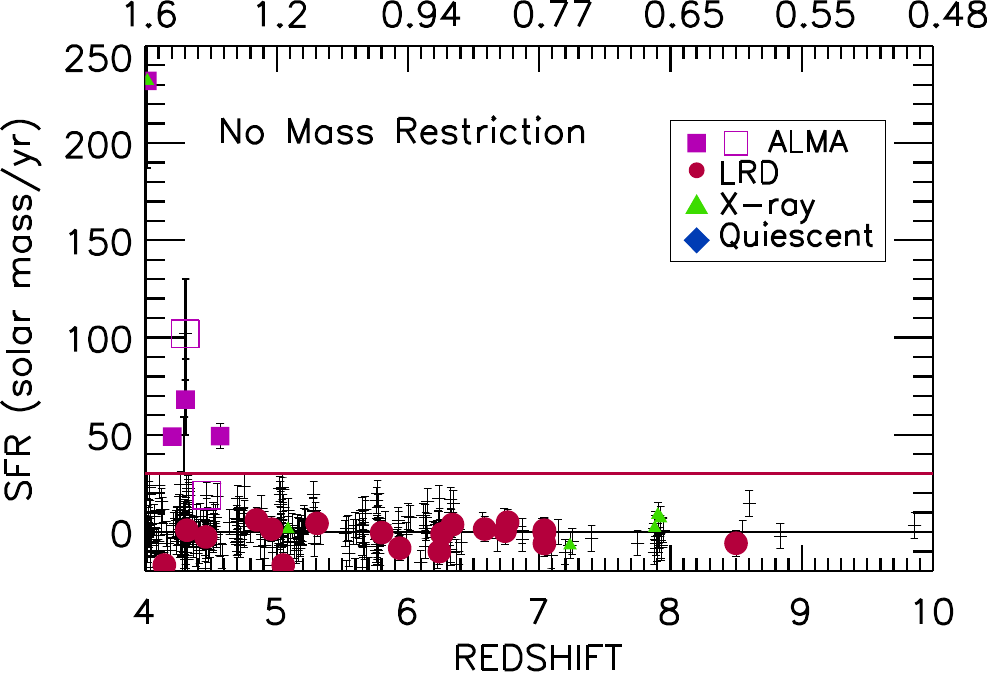}
\caption{
SFR vs. redshift for our JWST sample with (top) $z=1.5-5.5$
and $M_\ast=10^{8}$--$10^{10}$\md, 
(middle) $z=1.5-5.5$ and $M_\ast>10^{10}$\md,
and (bottom) $z>4$ without any mass restriction.
Along the top of each panel is the age in Gyr.
The red line shows SFR $=30$\,\sfr. 
The symbols are ALMA detections---purple squares (solid for sources
with $>4.5\sigma$ detections); open for $>3\sigma$);  
published LRDs---red circles;
quiescent galaxies satisfying the strict \citet{williams09}
selection---blue diamonds; bright X-ray sources---green triangles;
remaining sources---plus signs.
Note that none of the LRDs or quiescent galaxies are detected in the ALMA data.
}
\label{sfr_history}
\end{figure}

In Figure~\ref{sfr_history}, we show SFR versus redshift in
two mass intervals: $M_\ast=10^{8}$--$10^{10}$\md\ (top)
and $M_\ast>10^{10}$\md\ (middle).
Nearly all of the powerfully star-forming galaxies (above the red line),
and all of the $z>3$ powerfully star-forming galaxies,
have $M_\ast>10^{10}$\md;
none of the 685 galaxies with $M_\ast=10^{8}$--$10^{10}$\md\
and $z>3$ have a dust obscured SFR $>30$\,\sfr.

\begin{figure}
\includegraphics[width=8cm,angle=0]{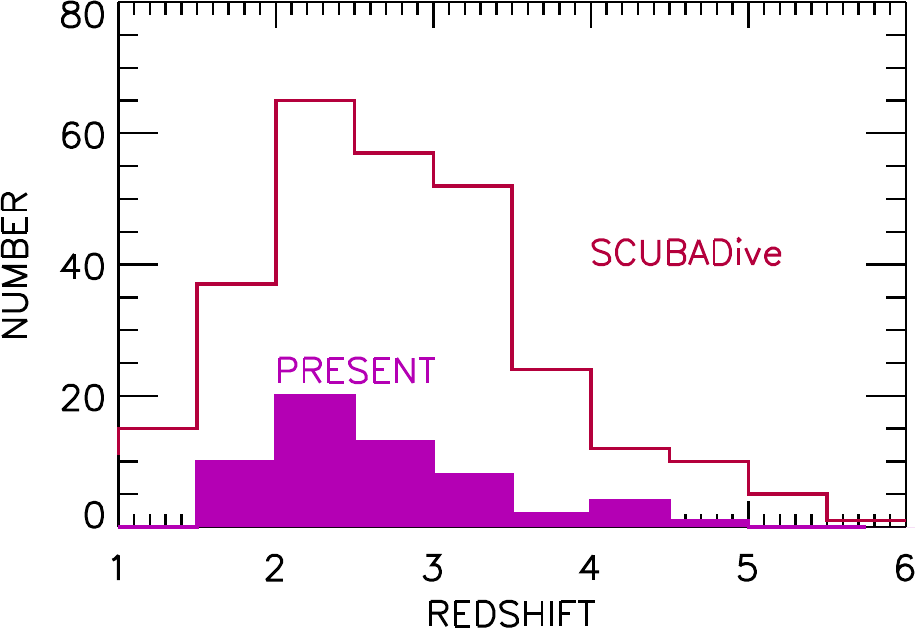}
\caption{
Comparison of the redshift distribution of our JWST sample with 
SFR $>30$\,\sfr (purple shading) with the 
redshift distribution of the SCUBADive sample of \cite{mckinney25}
(289 DSFGs). 
\label{zdist}
}
\end{figure}

Given that we cannot accurately estimate the rest-frame $J$-band fluxes 
in the JWST data at $z>5.5$, we can no longer sort by stellar mass at these redshifts.
However, we can still determine the SFRs based on their submillimeter
fluxes, which are completely independent of the mass determination.
In Figure~\ref{sfr_history} (bottom), we show the extension to these
higher redshifts for all galaxies, regardless of mass.
LRDs become more common at these redshifts, 
but none are detected in the ALMA data.
The mean of the 16 LRDs at z>4 is $-2.2 \pm 1.6$\,\sfr.  
We see that none of the galaxies beyond $z=4.57$ are detected at the
$>3\sigma$ level in the ALMA images, and all SFRs are $<30$\,\sfr.
This means that, in this sample, we are
seeing no significant star formation beyond $z=5$.

This is consistent with previous studies. Though
larger (and brighter) samples of DSFGs do show
small numbers of galaxies at the higher redshifts, they still show a rapid fall off. 
In Figure~\ref{zdist}, we show the redshift distribution for our JWST sample 
with SFR $>30$\,\sfr compared with that 
of SCUBADive (\citealt{mckinney25}).
SCUBADive is a large sample of 289 DSFGs. 
Although it is significantly brighter than our sample, 
it has a nearly identical redshift distribution, as do the samples of \citet{birkin21}, 
AS2COSMOS (\citealt{simpson20}), and AS2UDS (\citealt{dudzeviciute20}). 
The SCUBADive sample contains three spectroscopically identified galaxies 
at $z=5$--6, and a further three with photometric redshifts in this interval.
This 2\% rate is broadly consistent with the present data: for zero DSFGs at $z=5$--6, 
we have a $1\sigma$ upper limit of 1.8 \citep{gehrels86}
and a total sample of 68, giving an upper limit rate of 2.6\%.
However, it should be emphasized that the number of outliers in the photometric 
redshifts is substantial (around 20\%), which can result in contamination of the high-redshift tail 
\citep{mckay26}.

\citet{mckay26} used a near-complete spectroscopic sample of the SUPER GOODS 
sources from \citet{cowie18} to obtain a $1\sigma$ upper limit rate of 1.8\% at $z=5$--6,
which is again consistent with the present data.

\begin{figure}
\includegraphics[width=8cm,angle=0]{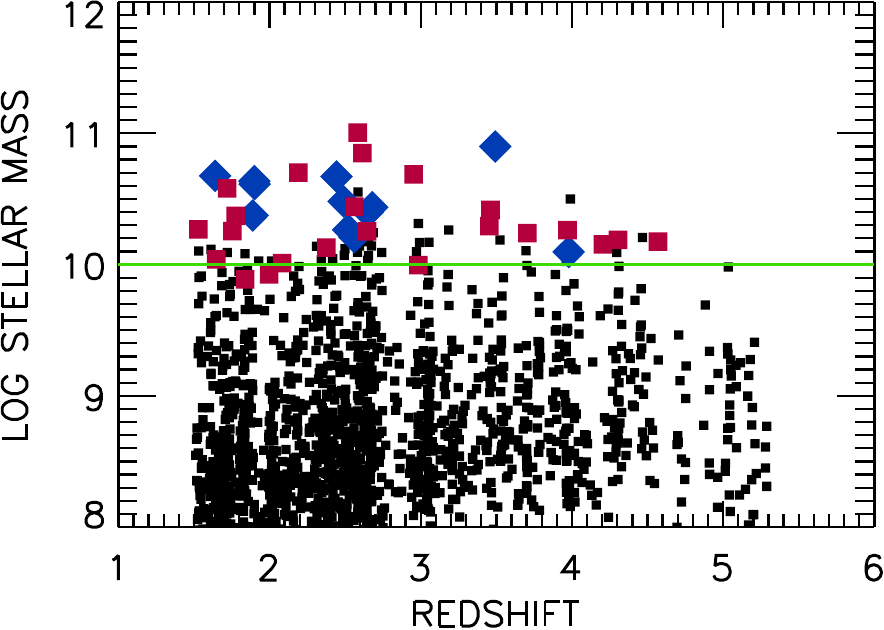}
\caption{Stellar mass vs. redshift for our JWST sample in the
redshift interval $z=1.5$--5.5.
The powerfully star-forming galaxies ($>30$\,\sfr)
(red squares) and quiescent galaxies (blue diamonds)
are mainly found in massive galaxies ($M_\ast>10^{10}$\md) 
(above the green line). 
\label{mass_zp_plot}
}
\end{figure}

In Figure~\ref{mass_zp_plot}, we show the evolution of the galaxy masses. 
We choose to focus on the more directly measured
quantities (stellar mass and SFRs) rather than more uncertain sSFRs. 
A discussion of these may be found in \cite{mckay25}.
Massive ($M_\ast>10$\md; above the green line) galaxies
become rare at $z>5$, paralleling
the fall off in the number of detected DSFGs. We note that
given the timescales, we can only marginally form $M_\ast>10^{10}$\md\ 
galaxies by $z=5$. For example, a galaxy that has a constant SFR = 25\,\sfr\
from $z=10$ to $z=5$ would form a stellar mass of $M_\ast=1.8\times10^{10}$\md.

Relative to the powerfully star-forming galaxies (red squares),
massive quiescent galaxies (blue diamonds) are a small fraction of the massive galaxies.
Of the two quiescent galaxies identified in the strict selection region at
$z>3$ (see Figure~\ref{fig:color_select}), the highest redshift is the $z=3.97$ source
at RA: $0^h$  $14^m$  $15.06^s$, Decl: $-30^\circ$  $23'$   $27\farcs4$.
The colors and spectrum of this source have been analyzed by \cite{setton24}, who
concluded that half the stellar mass had formed by $z=7.5$. If the source formed
stars at a constant rate between $z=10$ and $z=7.5$, then it would require
SFR = 28\,\sfr. We do not see such sources in the present
data at these high redshifts (see Figure~\ref{sfr_history}), but this could be a consequence 
of the field size in the source plane. However, it is also possible that few quiescent galaxies at $z\sim3-4$
require high-redshift star formation beyond $z=5$ and that the majority of the quiescent 
galaxies correspond to the star-forming galaxies seen in the $z=4-5$ interval in Figure~\ref{sfr_history}.
For example, \cite{nanayakkara24} found that only two out of 12 $z=3$ quiescent galaxies
in the ZFOURGE sample required large amounts of star formation at $z>5$ (see also
\citealt{nanayakkara25}).

\section{Summary}
In this paper, we used the rich dataset on the massive lensing cluster field A2744
to develop a new method for selecting DSFGs at $1.5<z<5.5$. 
The data we used in our analysis include new ALMA \alma\ observations and
DUALZ ALMA 1.2\,mm observations,
which cover much of the JWST UNCOVER area; SCUBA-2 850\,$\mu$m data;
Chandra X-ray data; and extensive JWST spectroscopic and photometric redshifts.
With the addition of the redshift information, we were able to develop a
rest-frame red color selection, which has the advantage over an observed-frame red
color selection of making it possible to separate out the different red populations,
to map and compare them as a function of redshift, and to determine galaxy properties, 
like NIR luminosity/stellar mass.

We found that a rest-frame red color selection of $f_{J}/f_{V} > 3$ and a high
NIR luminosity criterion of $L_\nu(1.2\,\mu$m)$>3.6\times10^{29}$\,erg\,s$^{-1}$\,Hz$^{-1}$
are very effective in selecting DSFGs, though at $z \lesssim 3$-4, 
we begin to see a small number of quiescent galaxies also populate this region. 
Using the strict selection of \cite{williams09}, we find two
quiescent galaxies in the redshift interval $z=3$--4
and zero above $z=4$. Using the padded selection, we
could add a further two possible quiescent sources
in the $z=3$--4 range.
Fortunately, these can be effectively separated from the DSFGs by their distinct location in 
the rest-frame $U-V$ versus $V-J$ diagram.

We do not find any low-mass ($M_\ast<10^{10}$\md)
ALMA or quiescent sources above $z=3$, though five
low-mass quiescent galaxies and one low-mass ALMA source
are seen in the redshift interval $z=2$--3.
When we consider powerfully star-forming galaxies (SFR $>30$\,\sfr), nearly all---and all
15 beyond $z=3$---are massive galaxies.

Our $f_{J}/f_{V} > 3$ selection also identifies a population of compact, less luminous red sources, 
most of which have already been identified as LRDs. 
The most luminous LRDs contaminate the quiescent galaxy region in the
rest-frame $U-V$ versus $V-J$ diagram.
We do not detect any of the LRDs or quiescent galaxies at the $>3\sigma$
level in the ALMA images. 
We also do not strongly detect in X-rays any of the LRDs in the field,
consistent with previous results. 

By applying this new selection technique, we were able to map the evolution of these different 
galaxy populations as a function of redshift. We found that massive galaxies become rare
at $z>5$, which parallels the fall off in the number of detected DSFGs, while LRDs are
more common at these high redshifts.

\begin{acknowledgements}
{
We wish to thank the anonymous referee for helpful suggestions that contributed substantially to improving 
the manuscript.
We gratefully acknowledge support for this research from the University of Wisconsin-Madison,
Office of the Vice Chancellor for Research with funding from
the Wisconsin Alumni Research Foundation (A.~J.~B.), 
NASA grant 80NSSC22K0483 (L.~L.~C.), and
the William F. Vilas Estate and the North American
ALMA Science Center through the ALMA Ambassadors program (S.~J.~M.).
F.~E.~B. acknowledges support from ANID-Chile BASAL CATA FB210003 and FONDECYT Regular 1241005.

The National Radio Astronomy Observatory is a facility of the National Science Foundation operated under cooperative 
agreement by Associated Universities, Inc.
This paper makes use of the following ALMA data: 
ADS/JAO.ALMA\#2013.1.00999.S, \\ 
ADS/JAO.ALMA\#2015.1.01425.S, \\ 
ADS/JAO.ALMA\#2017.1.01219.S, \\ 
ADS/JAO.ALMA\#2018.1.00035.L, \\ 
ADS/JAO.ALMA\#2021.1.00024.S,\\ 
ADS/JAO.ALMA\#2022.1.00073.S, \\ 
and ADS/JAO.ALMA\#2023.1.00468.S. \\ 
ALMA is a partnership of ESO (representing its member states), NSF (USA), and NINS (Japan), together with NRC (Canada), MOST and ASIAA (Taiwan), and KASI (Republic of Korea), in cooperation with the Republic of Chile. The Joint ALMA Observatory is operated by ESO, AUI/NRAO, and NAOJ.

The James Clerk Maxwell Telescope is operated by the East Asian Observatory on behalf of The National Astronomical Observatory of Japan, Academia Sinica Institute of Astronomy and Astrophysics, the Korea Astronomy and Space Science Institute, the National Astronomical Observatories of China and the Chinese Academy of Sciences (grant No.~XDB09000000), with additional funding support from the Science and Technology Facilities Council of the United Kingdom and participating universities 
in the United Kingdom and Canada. 

We wish to recognize and acknowledge 
the very significant cultural role and reverence that the summit of Maunakea has always had within the indigenous Hawaiian community. We are most fortunate to have the opportunity to conduct observations from this mountain.
}
\end{acknowledgements}

\facilities{ALMA, JWST, JCMT}

\footnotesize
\bibliography{red_bib}{}
\bibliographystyle{aasjournalv7}

\startlongtable
\centerwidetable
\begin{deluxetable*}{cccccccccccclll}
\renewcommand\baselinestretch{1.0}
\tablewidth{0pt}
\tablecaption{JWST Sample with ALMA $>4.5\sigma$ Detections\label{almatable}}
\scriptsize
\tablehead{
No. & Name & R.A. & Decl. &  \alma\  & 1.2\,mm & F444W &  $f_J/f_V$  & $z$  & $\mu$  & 0.5--1.2\,keV & 1.2--2\,keV & 2--8\,keV \\ 
& & J2000.0 & J2000.0 & (mJy) & (mJy) & ($\mu$Jy)  & &  &  & \multicolumn{3}{c}{($10^{-17}$\,erg\,s$^{-1}$)} \\
(1) & (2) & (3) & (4) & (5) & (6) & (7) & (8) & (9) & (10) & (11) & (12) & (13)
}
\startdata
   1 &  A001411-302317 & 3.5474167 & -30.388277 & 4.70(0.19) & 2.81(0.07) & 20.86 & 9.00 & 2.05 & 1.61 & 10 & 5 & 12\cr
   2 &  A001408-302137 & 3.5362918 & -30.360334 & 6.71(0.37) & 2.67(0.08) & 42.85 & 7.13 & 1.891 & 1.58 & 0 & 0 & 0\cr
   3 &  A001432-302615 & 3.6342499 & -30.437721 & 3.73(0.19) & \nodata & 6.107 & 10.7 & 1.52 & 1.27 & 8 & 3 & 8\cr
   4 &  A001411-302108 & 3.5491667 & -30.352251 & 3.93(0.20) & 1.87(0.07) & 17.92 & 5.09 & 2.489 & 1.47 & 35 & 38 & 189\cr
   5 &  A001418-302527 & 3.5755835 & -30.424362 & 6.13(0.41) & 3.01(0.12) & 4.968 & 8.45 & 4.014 & 1.63 & 6 & 12 & 32\cr
   6 &  A001419-302307 & 3.5825000 & -30.385473 & 2.98(0.22) & 1.49(0.06) & 7.180 & 4.10 & 3.056 & 4.61 & 4 & 5 & 7\cr
   7 &  A001418-302447 & 3.5760000 & -30.413195 & 7.43(0.50) & 3.04(0.11) & 3.638 & 9.21 & 2.03 & 1.82 & 10 & 0 & 10\cr
   8 &  A001428-302207 & 3.6172502 & -30.368807 & 4.49(0.33) & 1.97(0.09) & 31.59 & 2.63 & 0.52 & 1.15 & 60 & 51 & 77\cr
   9 &  A001417-302300 & 3.5732501 & -30.383501 & 2.21(0.20) & 0.91(0.06) & ? & ? & ? & ? & 0 & 2 & 13\cr
  10 &  A001430-302339 & 3.6276667 & -30.394278 & 1.96(0.19) & 1.05(0.08) & 2.084 & 5.40 & 3.36 & 1.42 & 0 & 2 & 4\cr
  11 &  A001403-302251 & 3.5126665 & -30.380972 & 1.76(0.20) & 0.71(0.14) & 6.501 & 4.56 & 2.56 & 1.36 & 24 & 3 & 0\cr
  12 &  A001420-301959 & 3.5847917 & -30.333166 & 1.76(0.17) & \nodata & 10.48 & 5.52 & 2.32 & 1.34 & 0 & 4 & 0\cr
  13 &  A001413-302117 & 3.5582917 & -30.354944 & 2.11(0.23) & 1.22(0.09) & 2.626 & 4.75 & 2.00 & 1.56 & 0 & 0 & 3\cr
  14 &  A001419-302242 & 3.5797083 & -30.378416 & 2.45(0.26) & 0.93(0.07) & 21.18 & 7.54 & 2.31 & 2.68 & 0 & 0 & 0\cr
  15 &  A001420-302254 & 3.5850000 & -30.381805 & 2.62(0.38) & 1.71(0.12) & 24.05 & 4.24 & 3.058 & 3.13 & 12 & 0 & 22\cr
  16 &  A001424-302145 & 3.6005836 & -30.362720 & 1.60(0.34) & 0.75(0.05) & 8.031 & 9.70 & 2.496 & 1.50 & 0 & 3 & 25\cr
  17 &  A001422-302123 & 3.5938752 & -30.356638 & 1.07(0.21) & 1.54(0.11) & 107.8 & 4.68 & 0.707 & 1.49 & 8 & 2 & 0\cr
  18 &  A001414-302503 & 3.5601251 & -30.417557 & 1.28(0.22) & \nodata & 24.79 & 5.45 & 1.48 & 1.34 & 7 & 5 & 68\cr
  19 &  A001416-302341 & 3.5668333 & -30.394890 & 1.63(0.34) & 0.70(0.05) & 9.134 & 6.28 & 3.054 & 2.12 & 31 & 4 & 47\cr
  20 &  A001415-302039 & 3.5641668 & -30.344444 & 2.17(0.29) & 0.97(0.09) & 6.732 & 2.94 & 0.37 & 1.00 & 2 & 0 & 8\cr
  21 &  A001417-302258 & 3.5720000 & -30.382944 & 1.91(0.47) & 0.68(0.05) & 45.83 & 8.83 & 1.672 & 2.79 & 0 & 10 & 21\cr
  22 &  A001409-302136 & 3.5391667 & -30.360250 & 1.48(0.40) & 0.59(0.05) & 127.6 & 10.6 & 1.33 & 1.54 & 278 & 351 & 1429\cr
  23 &  A001423-302135 & 3.5990419 & -30.359751 & 1.29(0.25) & 0.68(0.05) & 62.46 & 4.82 & 1.360 & 1.41 & 4 & 2 & 10\cr
  24 &  A001420-302530 & 3.5863748 & -30.425056 & 2.33(0.42) & 0.79(0.05) & 3.106 & 5.12 & 2.974 & 1.78 & 3 & 0 & 0\cr
  25 &  A001407-302140 & 3.5312083 & -30.361307 & 0.89(0.23) & 0.50(0.05) & 31.63 & 9.81 & 1.28 & 1.54 & 10 & 8 & 23\cr
  26 &  A001413-302219 & 3.5543332 & -30.371973 & 2.37(0.44) & 0.64(0.05) & 22.31 & 6.04 & 2.387 & 3.47 & 4 & 1 & 2\cr
  27 &  A001405-302138 & 3.5211248 & -30.360666 & 2.37(0.42) & 0.54(0.06) & 62.36 & 5.25 & 1.529 & 1.52 & 0 & 0 & 0\cr
  28 &  A001415-302507 & 3.5632501 & -30.418694 & 1.22(0.41) & \nodata & 3.893 & 5.17 & 2.980 & 1.36 & 0 & 0 & 1\cr
  29 &  A001405-302217 & 3.5237501 & -30.371445 & 1.01(0.21) & 0.33(0.05) & 8.687 & 4.62 & 2.387 & 2.02 & 0 & 0 & 17\cr
  30 &  A001414-302239 & 3.5599167 & -30.377777 & 1.23(0.36) & 0.28(0.05) & ? & ? & ? & ? & 30 & 28 & 1150\cr
  31 &  A001420-302523 & 3.5872500 & -30.423056 & 0.86(0.16) & \nodata & 5.679 & 3.19 & 2.46 & 1.90 & 0 & 0 & 1\cr
  32 &  A001413-302238 & 3.5579166 & -30.377279 & 0.88(0.24) & 0.37(0.05) & 21.62 & 2.21 & 2.578 & 9.03 & 0 & 3 & 0\cr
  33 &  A001424-302546 & 3.6036251 & -30.429556 & \nodata & 0.27(0.05) & 25.29 & 4.82 & 1.52 & 1.58 & 0 & 0 & 1\cr
  34 &  A001425-302505 & 3.6056249 & -30.418056 & \nodata & 0.27(0.05) & ? & ? & ? & ? & 0 & 0 & 2\cr
  35 &  A001426-302452 & 3.6085835 & -30.414583 & \nodata & 0.54(0.05) & 34.98 & 3.14 & 2.613 & 2.17 & 0 & 2 & 20\cr
  36 &  A001415-302443 & 3.5650415 & -30.412193 & \nodata & 0.27(0.05) & 13.89 & 4.48 & 1.27 & 1.43 & 0 & 3 & 2\cr
  37 &  A001416-302410 & 3.5690000 & -30.402805 & \nodata & 0.29(0.05) & 45.37 & 7.40 & 2.583 & 1.81 & 0 & 0 & 8\cr
  38 &  A001417-302345 & 3.5723751 & -30.395945 & \nodata & 0.39(0.05) & 6.256 & 3.18 & 0.75 & 1.84 & 24 & 6 & 0\cr
  39 &  A001424-302346 & 3.6005418 & -30.396168 & \nodata & 0.17(0.03) & 32.45 & 4.07 & 0.945 & 1.93 & 8 & 0 & 7\cr
  40 &  A001428-302342 & 3.6175418 & -30.395000 & \nodata & 0.28(0.05) & 1.575 & 1.66 & 4.569 & 1.61 & 0 & 0 & 0\cr
  41 &  A001431-302337 & 3.6326668 & -30.393639 & \nodata & 0.60(0.05) & 1.540 & 4.09 & 3.970 & 1.38 & 0 & 0 & 4\cr
  42 &  A001429-302335 & 3.6214998 & -30.393112 & \nodata & 0.56(0.05) & 3.544 & 3.03 & 3.45 & 1.51 & 0 & 0 & 0\cr
  43 &  A001419-302326 & 3.5809999 & -30.390751 & 0.47(0.14) & 0.28(0.05) & ? & ? & ? & ? & 5 & 6 & 0\cr
  44 &  A001407-302314 & 3.5327084 & -30.387333 & \nodata & 0.37(0.05) & 24.35 & 2.38 & 2.190 & 1.43 & 7 & 2 & 10\cr
  45 &  A001404-302310 & 3.5190415 & -30.386332 & \nodata & 1.50(0.20) & ? & ? & ? & ? & 12 & 2 & 1\cr
  46 &  A001432-302256 & 3.6357501 & -30.382389 & \nodata & 0.59(0.05) & 8.548 & 6.97 & 2.35 & 1.30 & 12 & 1 & 0\cr
  47 &  A001422-302249 & 3.5920832 & -30.380499 & \nodata & 0.53(0.05) & 9.189 & 14.6 & 2.645 & 2.24 & 18 & 35 & 63\cr
  48 &  A001419-302248 & 3.5812917 & -30.380249 & \nodata & 0.33(0.05) & 3.620 & 6.77 & 3.475 & 3.13 & 23 & 0 & 2\cr
  49 &  A001419-302237 & 3.5823753 & -30.377167 & \nodata & 0.29(0.05) & ? & ? & ? & ? & 8 & 1 & 7\cr
  50 &  A001423-302228 & 3.5995834 & -30.374695 & \nodata & 0.38(0.05) & 8.314 & 6.57 & 2.083 & 1.64 & 0 & 3 & 2\cr
  51 &  A001402-302231 & 3.5103331 & -30.375444 & \nodata & 0.35(0.06) & 14.54 & 3.28 & 0.94 & 1.30 & 0 & 0 & 16\cr
  52 &  A001415-302203 & 3.5637918 & -30.367611 & \nodata & 0.26(0.04) & 78.63 & 2.86 & 1.324 & 1.94 & 5 & 6 & 0\cr
  53 &  A001409-302156 & 3.5375836 & -30.365639 & \nodata & 0.34(0.05) & 41.23 & 4.46 & 1.35 & 1.90 & 0 & 6 & 4\cr
  54 &  A001405-302133 & 3.5237501 & -30.359222 & \nodata & 0.29(0.05) & 7.675 & 3.28 & 2.387 & 1.54 & 0 & 0 & 0\cr
  55 &  A001422-302122 & 3.5926247 & -30.356140 & \nodata & 0.44(0.05) & 10.75 & 4.36 & 2.384 & 1.58 & 0 & 0 & 0\cr
  56 &  A001411-302112 & 3.5463750 & -30.353472 & \nodata & 0.27(0.05) & 4.571 & 4.44 & 2.497 & 1.47 & 0 & 0 & 1\cr
  57 &  A001407-302108 & 3.5305417 & -30.352304 & 1.27(0.34) & 0.31(0.05) & 29.26 & 4.48 & 1.35 & 1.35 & 0 & 0 & 0\cr
  58 &  A001414-301937 & 3.5612917 & -30.326973 & \nodata & 0.21(0.04) & 1.046 & 4.68 & 4.20 & 1.31 & 0 & 0 & 16\cr
  59 &  A001413-302008 & 3.5555418 & -30.335638 & \nodata & 0.26(0.05) & 2.014 & 2.30 & 3.70 & 1.36 & 0 & 0 & 0\cr
  60 &  A001412-302038 & 3.5528333 & -30.344166 & \nodata & 0.44(0.05) & 31.96 & 3.27 & 1.71 & 1.37 & 4 & 0 & 1\cr
  61 &  A001414-301948 & 3.5610831 & -30.330057 & \nodata & 0.41(0.05) & 39.19 & 8.02 & 1.33 & 1.26 & 0 & 0 & 0\cr
  62 &  A001433-302252 & 3.6384580 & -30.381222 & \nodata & 0.30(0.06) & ? & ? & ? & ? & 12 & 0 & 17\cr
  63 &  A001413-302228 & 3.5582497 & -30.374445 & \nodata & 0.26(0.05) & 18.40 & 2.81 & 2.407 & 4.87 & 2 & 14 & 27\cr
  64 &  A001422-302206 & 3.5923750 & -30.368389 & \nodata & 0.26(0.05) & blank & ? & ? & ? & 0 & 0 & 2\cr
  65 &  A001413-301952 & 3.5580418 & -30.331112 & \nodata & 0.26(0.05) & 17.83 & 2.54 & 1.775 & 1.29 & 6 & 0 & 17\cr
  66 &  A001414-301951 & 3.5617917 & -30.330944 & \nodata & 0.28(0.05) & ? & ? & ? & ? & 0 & 0 & 12\cr
  67 &  A001416-301946 & 3.5685835 & -30.329639 & \nodata & 1.13(0.24) & blank & ? & ? & ? & 3 & 1 & 0\cr
  68 &  A001414-302458 & 3.5605416 & -30.416140 & 0.80(0.17) & \nodata & 24.32 & 2.17 & 2.290 & 1.39 & 11 & 15 & 176\cr
\enddata
\tablecomments{The fluxes are in the observed frame.
Sources that are blended/contaminated
by a foreground object in F444W are marked as `?' in the F444W, $f_J/f_V$, $z$, and magnification columns.  
$1\sigma$ uncertainties are given in parentheses for the 870\,$\mu$m and 1.2\,mm fluxes but not for the 
JWST F444W flux, since they are so small. However, they can be found in the \citet{weaver24} catalog. 
Only source~3 was not
included in their catalog, so we measured its flux using a $0\farcs5$ radius aperture on the 
\citet{bezanson24} F444W image.
The 68\% confidence errors on the magnification are all less than 25\%,
with only four greater than 10\%.
Two of the lowest S/N submillimeter sources have no F444W counterparts (sources 64 and 67) and are 
marked as `blank' in that column and as `?' in the $f_J/f_V$, $z$, and magnification columns.
}
\end{deluxetable*}

\end{document}